\documentclass[a4paper]{article}

\usepackage[utf8]{inputenc}

\usepackage{graphicx}
\usepackage{subcaption}
\usepackage{mathtools}
\usepackage{wrapfig}
\usepackage{graphicx}
\usepackage{subcaption}
\usepackage{mathtools}
\usepackage{wrapfig}
\usepackage{float}
\usepackage{gensymb}
\usepackage{caption}
\usepackage{geometry}
\usepackage{color,soul}
\usepackage{slashbox}
\usepackage{amsfonts}
\usepackage{natbib}

\makeatletter
\newcommand*{\rom}[1]{\expandafter\@slowromancap\romannumeral #1@}
\makeatother

\usepackage{listings}
\usepackage{color}

\definecolor{codegreen}{rgb}{0,0.6,0}
\definecolor{codegray}{rgb}{0.5,0.5,0.5}
\definecolor{codepurple}{rgb}{0.58,0,0.82}
\definecolor{backcolour}{rgb}{0.95,0.95,0.92}
 
\lstdefinestyle{mystyle}{
    backgroundcolor=\color{backcolour},   
    numberstyle=\tiny\color{codegray},
    basicstyle=\footnotesize,
    breakatwhitespace=false,         
    breaklines=true,                 
    captionpos=b,                    
    keepspaces=true,                 
    numbers=left,                    
    numbersep=5pt,                  
    showspaces=false,                
    showstringspaces=false,
    showtabs=false,                  
    tabsize=2
}

\makeatletter
\renewcommand\paragraph{\@startsection{paragraph}{4}{\z@}%
	{-2.5ex\@plus -1ex \@minus -.25ex}%
	{1.25ex \@plus .25ex}%
	{\normalfont\normalsize\bfseries}}
\makeatother
\usepackage{pdfpages}

\usepackage{lipsum}

\newcommand\blfootnote[1]{%
  \begingroup
  \renewcommand\thefootnote{}\footnote{#1}%
  \addtocounter{footnote}{-1}%
  \endgroup
}

\begin{document} 
\title{THE SENSITIVITY OF TRIVARIATE GRANGER CAUSALITY TO TEST CRITERIA AND DATA ERRORS} 
\author{LEO CARLOS-SANDBERG AND CHRISTOPHER D. CLACK\\
Department of Computer Science, University College London\\
leo.carlos-sandberg.16@ucl.ac.uk clack@cs.ucl.ac.uk\\
} 
\maketitle 

\begin{abstract}
\noindent {\it Trivariate Granger causality analysis seeks to distinguish between ``true" causality and ``spurious" causality results from the topology of the system. However, this analysis is sensitive both to the choice of test criteria and the presence of noise, and this can lead to incorrect inference of causality: either to infer causality that does not exist (spurious causality), or to fail to infer causality that does exist (unidentified causality). Here we analyse the effects of the choice of test criteria and the presence of noise and give general conditions under which incorrect inference is likely to occur. By studying the test criteria (likelihood ratio, Lagrange multiplier, Rao efficient scoring and Wald), we demonstrate that Rao efficient scoring and Wald tests are statistically indistinguishable and that for small sample sizes they offer a the lowest  likelihood of spurious causality, with the likelihood ratio test offering the lowest likelihood of unidentified causality. We also show the sample size at which convergence between these tests occurs. We also give empirical results for intrinsic noise (in a variable) and extrinsic noise (between an variable and a observer), with a varying signal-to-noise ratio for each variable, showing that for intrinsic noise a strong dependence on the signal-to-noise ratio of the last variable exists, and for extrinsic noise no dependence the true topology exists.  \blfootnote{\hspace{-6pt}\scriptsize\textcopyright \hspace{0.5mm} Leo Carlos-Sandberg and Christopher D. Clack 2019. \hfill\break\indent This work is licensed under a {Creative Commons Attribution 4.0 International License (CC  BY):} \hfill\break\indent https://creativecommons.org/licenses/by/4.0/  \hfill\break\indent Provided you adhere to the CC BY license, including as to attribution, you are free to copy and redistribute this \hfill\break\indent  work in any medium or format and remix, transform, and build upon the work for any purpose, even commercially.    \vspace{1mm} }
\vfill
}
\end{abstract}

\section{Introduction} \label{sec:introduction}

The determination of a causal topology among a collection of variables is an important issue that must be concluded in a number of scenarios, including construction of econometric models. Failure to correctly identify these topologies may lead to erroneous results and misleading interpretations. 

Granger causality has been an important tool for investigating causal relationships within time series data since its inception in 1969 \citep{Granger_1969}. Its definition of causality allows for relationships between time series to be investigated without any more information than the time series data and without the need for a perturbation to the time series (a more thorough description is given by \cite{Ding_etal2006}). 

The classical form of Granger causality is designed to investigate strictly bivariate systems, yet in the presence of one or more {\em unseen} variables bivariate Granger causality can lead to ``spurious causality'', where a causal link is found that does not exist in reality (with the opposite being ``unidentified causality'' where a causal link that exists is not found) \citep{Ding_etal2006}.  

Systems containing three known variables are referred to as trivariate; these trivariate systems require a slightly adapted form of Granger causality to account for the possibility that pair-wise analysis might similarly produce spurious causality, as discussed by \cite{Ding_etal2006}. 

While bivariate Granger causality has seen significant use \citep{Eichler_2012}, its extension to collections of variables through the use of trivariate Granger causality has had little investigation, especially when relating to its own limitations. In this paper we will outline potential causes of spurious and unidentified causality which should be considered when using the trivariate Granger method. Knowledge of these issues will allow better understanding of potential bias contained with the results gained from the use of this method allowing for improved interpretations. 

To further the understanding of the limitations of the trivariate Granger causality method this paper borrows on work from two papers (\cite{Taylor_1989} and \cite{Anderson_2018}) focusing on the analysis of spurious causality occurring in the bivariate Granger method.  

Test criteria have been investigated previously for the bivariate case by \cite{Taylor_1989}, who notes that a number of statistical tests exists for Granger causality that can lead to contradicting results (for example \cite{Sims_1972},  \cite{Sargent_1976}, and \cite{PierceHaugh_1977}). \cite{TiaoBox_1981} suggest a more appropriate method, which has since found prominent use in analysis of Granger causality. Though this method is widely used, there are four alternative approaches to it, since its testing of the null hypothesis can be performed using either the likelihood ratio (LR), Lagrange multiplier (LM), Rao efficient scoring (R), or Wald (W) testing criteria. Inspired 
by \cite{Taylor_1989}, we will show results comparing the different test criteria, looking at the probability of spurious or unidentified causality occurring.
 
\cite{Anderson_2018} have investigated the effect of ``errors-in-variable'' for a bivariate set-up; this can be viewed as an investigation of how a noise term present within a time series can lead to spurious causality. Here we expand on this work, focusing on a trivariate system, to look at both (i) noise terms present within a variable (which we will refer to as an ``intrinsic noise'') and (ii) noise terms present when an external observer accesses a variable's data (which we will refer to as an ``extrinsic noise''). A selection of example cases are chosen to represent the probabilities of spurious or unidentified causality occurring for a Granger causality test on a trivariate system when a selection of combinations of intrinsic and extrinsic noise terms are present.

This paper is organised as follows. In Section~\ref{sec:granger_causality} Granger causality is more fully defined and the current approach for implementing bivariate and trivariate versions of the tests are described. Section~\ref{sec:comparison_of_test_criteria} details an investigation into the effect of test criteria on the probability of spurious and unidentified causality, with Section~\ref{sec:tc_resutls} presenting an analysis of this investigation. Section~\ref{sec:sensitivity_to_noise} details an experiment investigating and analysing the effects of both intrinsic and extrinsic noise terms on the rate of spurious and unidentified causality. Section~\ref{sec:conclusion} concludes the paper.

\section{Granger Causality} \label{sec:granger_causality}

Granger provides a testable definition of causality, where one variable is considered to be caused by another if the information contained in the first improves the prediction of the second \citep{Granger_1969}. In this paper we use vector autoregression (VAR) models to express how these variables interact.

A bivariate Granger causality test (to see if variable $X$ causes variable $Y$) is undertaken by first setting up an equation to model $Y$, using both $Y$ and $X$ as predictors. This is referred to as the unrestricted model and is given as
\begin{equation}\label{eq:biurm}
y_{t} = \sum_{i=1}^{i_{max}}\alpha_{i}y_{t-i} + \sum_{j=1}^{j_{max}}\beta_{j}x_{t-j}+\varepsilon_{t}, 
\end{equation}  
where the values for $\alpha$ and $\beta$ are determined by an ordinary least squares fit to observed data \citep{Siggiridou_2016} and $i_{max}$ and $j_{max}$ are the ``look back'' of the time series. The values of $\varepsilon$ represent the residual error of the model (the difference between the predicted value of $y_{t}$ and the actual value of $y_{t}$); this is not the same as the intrinsic and extrinsic noise which will be discussed later ($\varepsilon$ is a modelling error, whereas the intrinsic noise is an noise in the creation of the time series, and the extrinsic noise is a noise in the observation of the time series). The unrestricted model, in Eq.~\ref{eq:biurm}, is then compared with a restricted model (where $X$ is not used as a predictor for $Y$). The restricted model is given as 
\begin{equation}\label{eq:birm}
y_{t} = \sum_{i=1}^{i_{max}}\alpha_{i}y_{t-i} + \varepsilon_{t}, 
\end{equation} 
An hypothesis test (discussed in greater detail by \cite{Uriel_2013}) is then performed using the following null hypothesis
\begin{equation*}
H_{0}: \beta_{1} = \beta_{2} = ... = \beta_{j_{max}} = 0 
\end{equation*}
With the alternative hypothesis $H_{1}$ being that the null hypothesis $H_{0}$ is not true. If the null hypothesis $H_{0}$ is not rejected then it is said that there exists no Granger causality between $X$ and $Y$, if the null hypothesis is rejected it is said that there exists Granger causality between $X$ and $Y$. 

\begin{figure}[H]
	\centering
	\includegraphics[scale=0.5]{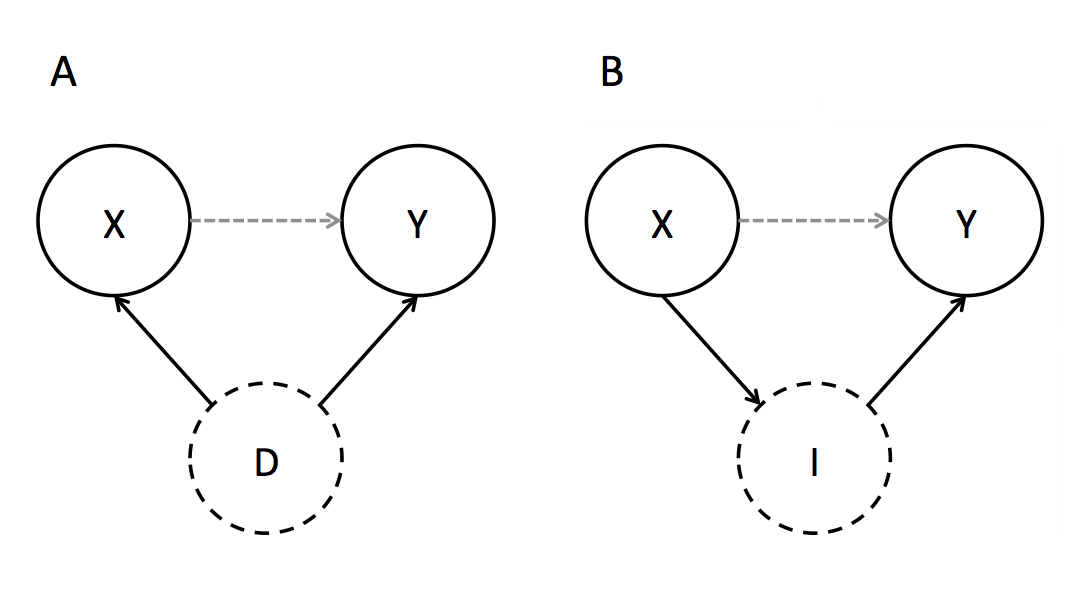}
	\caption{\it A. A driver case, where an unknown variable, $D$ (shown as a dashed circle), is causing both $X$ and $Y$, leading to a spurious link between them (shown as a dashed arrow). B. A indirect case, where $X$ causes an unknown variable $I$ (shown as a dashed circle) that in turn causes $Y$, leading to a spurious link between $X$ and $Y$ (shown as a dashed arrow). In B one may argue that the link between $X$ and $Y$ is genuine given that the effects of $X$ do effect $Y$ via $I$, however this approach can lead to important transformations to the data occurring within $I$ being missed, and hence unexpected behaviour being seen in the link between $X$ and $Y$.}  
	\label{fig:unknownvaribles}
\end{figure}

The topology of a bivariate system is simple, with the two variables being either causally related or not causally related (ignoring feedback loops). The presence of a third unknown variable interacting with this system can however lead to spurious causality between the two known variables being inferred. This spurious causality can occur in two types of topologies, either with the unknown variable acting as a driver or facilitating an indirect interaction, as shown in Figure~\ref{fig:unknownvaribles}. 

\begin{figure}[H]
	\centering
	\includegraphics[scale=0.5]{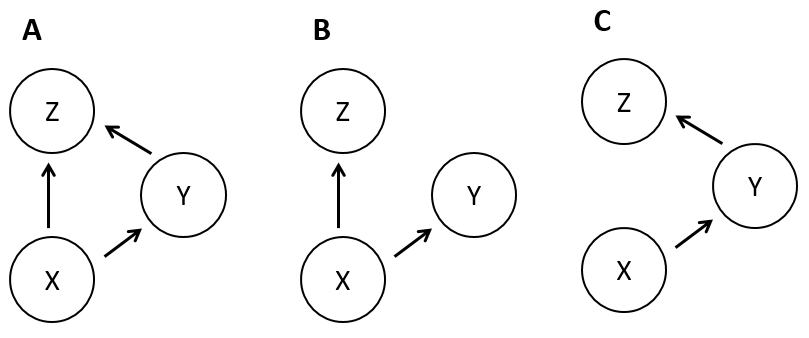}
	\caption{\it Trivariate topologies that under bivariate Granger causality tests may all wield results shown in A. A - complete topology. B - driver topology, where $X$ drives both $Y$ and Z, and bivariate analysis may find a link from $Y$ to $Z$. C - indirect topology, where $X$ causes $Z$ via an intermediary $Y$, and bivariate analysis may find a link from $X$ to $Z$.}  
	\label{fig:figexampletritopolgies}
\end{figure}

Having knowledge of this unknown variable does not immediately solve the problem of spurious causality, since a bivariate analysis of the system will still incorrectly identify a link. Due to this, when investigating trivariate systems with a bivariate approach, three types of topology will be indistinguishable from each other, shown in Figure~\ref{fig:figexampletritopolgies} (these are base topologies and more complex expressions can exists which embody the same issue). It is also worth noting that even for the complete case where the causality will be correctly inferred, the coefficients associated with the causal links are likely to be incorrect. 

In this paper we present results related to two of these topologies, the driver and indirect topologies (Figure\ref{fig:figexampletritopolgies} B, and C respectively), and the possible issues distinguishing between them. To allow the results between these topologies to be more readily compared, time delays will be set so that events in $Y$ occur one time step after those in $X$, and the events in $Z$ occur two time steps after those in $X$ (one time step after those in $Y$). This particular set up allows any $Z$ to be created with either the driver or indirect topology, allowing our results to not be influenced by any change in $Z$ that would occur by a switching between the topologies using a different time delay (e.g. if the time delay between $X$ and $Z$ was one time step then a driver topology would create a $Z$ time series one step ahead of that created in a indirect topology). 

To distinguish between these types of topologies a trivariate Granger causality test has to be employed, and this test is performed in two steps. First all possible bivariate links should be tested using the bivariate Granger causality test; however if this returns a complete topology, Figure\ref{fig:figexampletritopolgies} A, the topology may contain spurious links. Second, spurious links should be explored by running two tests, one to see if $X$ improves the prediction of $Z$ and one to see if $Y$ improves the prediction of $Z$. Using the former as an example, an unrestricted model would be created in the form of
\begin{equation}
z_{t} = \sum_{i=1}^{i_{max}}\alpha_{i}z_{t-i} + \sum_{j=1}^{j_{max}}\beta_{j}y_{t-j} + \sum_{k=1}^{k_{max}}\gamma_{j}x_{t-j} +\varepsilon_{t} \label{eq:unrestriex}
\end{equation}
and would be tested against the restricted model 
\begin{equation}
z_{t} = \sum_{i=1}^{i_{max}}\alpha_{i}z_{t-i} + \sum_{j=1}^{j_{max}}\beta_{j}y_{t-j} +\varepsilon_{t} \label{eq:unrestriex}
\end{equation}
These two models are then compared using the hypotheses
\begin{equation*}
H_{0}: \gamma_{1} = \gamma_{2} = ... = \gamma_{k_{max}} = 0 
\end{equation*}  
With the alternative hypothesis being that the null hypothesis is not true. If the null hypothesis is rejected then it is said that there is Granger causality directly between $X$ and $Z$ and if the null hypothesis is not rejected then it is said that the link found between $X$ and $Z$ in the bivariate analysis is spurious. This process is repeated for the link between $Y$ and $Z$, with the results producing a final topology shown in Figure~\ref{fig:figexampletritopolgies}.

\section{Comparison of Test Criteria}\label{sec:comparison_of_test_criteria}
In the decision of whether to accept or reject the null hypothesis during Granger analysis a p-value is used (calculated by a test criterion), if this value is greater than a set significance level ($\alpha$) then it is decided to reject the null hypothesis and hence infer a causal link. For trivariate Granger analysis a look-back of greater then one is normally required to capture information from each variable, hence the hypothesis test to calculate the p-value must support a look-back of greater then one (multiple hypothesis testing).

Following the work done by \cite{Taylor_1989} four classical test criteria appropriate for Granger analysis are stated: LR, W, LM and R (as described in the introduction). These criteria are described in detail by \cite{Judge_etal1980}, but a short intuitive description of how they are related to the unrestricted and restricted model is give here (by following \cite{Taylor_1989}). 

\begin{itemize}
	\item The LR test looks at the ratio of the likelihood function for the restricted model and for the unrestricted model. 
	\item The W test uses just the unrestricted model, it then evaluates the parameters to be restricted to create the restricted model (we use the Chi-squared distribution).
	\item The LM test uses only the restricted model and evaluates how relaxing the restrictions imposed on this model improves its fit.
	\item  The R test evaluates how far the slope vector of the restricted model deviates from the zero vector. 
\end{itemize}

Taylor states that in order to compare these test criteria one must consider both the computational efficiency and the power characteristics of each. It should be noted that the LM test can be shown to be dominated by the R test, due to them both having an equal power function \citep{Taylor_1989}, and since the LM test is computationally less efficient, the LM test will not be considered for the remainder of this paper. Therefore the set of test criteria being investigated is reduced to the LR, W, and R tests; furthermore we use a more recent version of the R test which has improved results for small data sets \citep{Hermes_2014}. These test criteria can be shown to produce statistics that have the same conditional asymptotic distributions, under an increasing number of data points. However for most real world examples the sample sizes will be relatively small, leading to deviations in the numerical values calculated by these criteria. These deviations in the calculated test statistic can lead to the test criteria inferring different causal topologies for the same significance level. The bivariate investigation into the small sample properties of these test criteria when applied to Granger causality, by \cite{Taylor_1989}, notes that conclusions drawn may not apply to trivariate systems, observing work by \cite{Sims_1980}, and \cite{Litterman_etal1985} which suggest the presence of a third variable may influence the bivariate results.   

When applying this investigation to the trivariate Granger case both the probability of spurious causality and the probability of unidentified causality will be important to note for the comparison between the criteria. 

\subsection{Test Functions} \label{sec:tc_testfunctions}

For this investigation synthetic test data was used, and for simplicity this data was generated using simple linear models, similar to those used in \cite{Granger_1969}. 

Test data was generated for two types of topologies, an indirect and a driver (shown in Figure~\ref{fig:figexampletritopolgies}) topology. For both these topologies results for the rate of occurrence of both spurious and unidentified causality can be calculated (results for the complete topology were not included since spurious causality can not occur in the base case). These rates were generated using a Monte Carlo experiment with $5,000$ iterations run with the following equations, 
\begin{align*}
&x_{t} = U(-2, 2)\\
&y_{t} = 0.3\times y_{t-1} + x_{t-1} + N(0, 0.1)\\
&z_{t} = 0.3\times z_{t-1} + x_{t-2} + N(0, 0.5)\\
&z_{t}^{'} = 0.3\times z_{t-1}^{'} + y_{t-1} + N(0, 0.5)\\
\end{align*}
where $U(a, b)$ is the uniform distribution, and $N(\mu, \sigma )$ is the normal distribution (used for the noise term). This creates four series $X$, $Y$, $Z$ and $Z^{'}$, where $Z$ is a driver topology and $Z^{'}$ is an indirect topology. 

The time series $Z$ is defined in terms of a two time step delay for the values of $X$; as previously explained, this allows for $Z$ and $Z^{'}$ to be indistinguishable (apart from possible noise terms) if they are created with the same $X$ and $Y$ time series. The noise term on both $Z$ and $Z^{'}$ was chosen to have a higher standard deviation then that of $Y$, to mask the $X$ and $Y$ signals and give a larger dispersion in the Granger results.

\subsection{Results and Analysis} \label{sec:tc_resutls}

\begin{figure}
\centering
\begin{subfigure}[b]{0.55\textwidth}
   \includegraphics[width=1\linewidth]{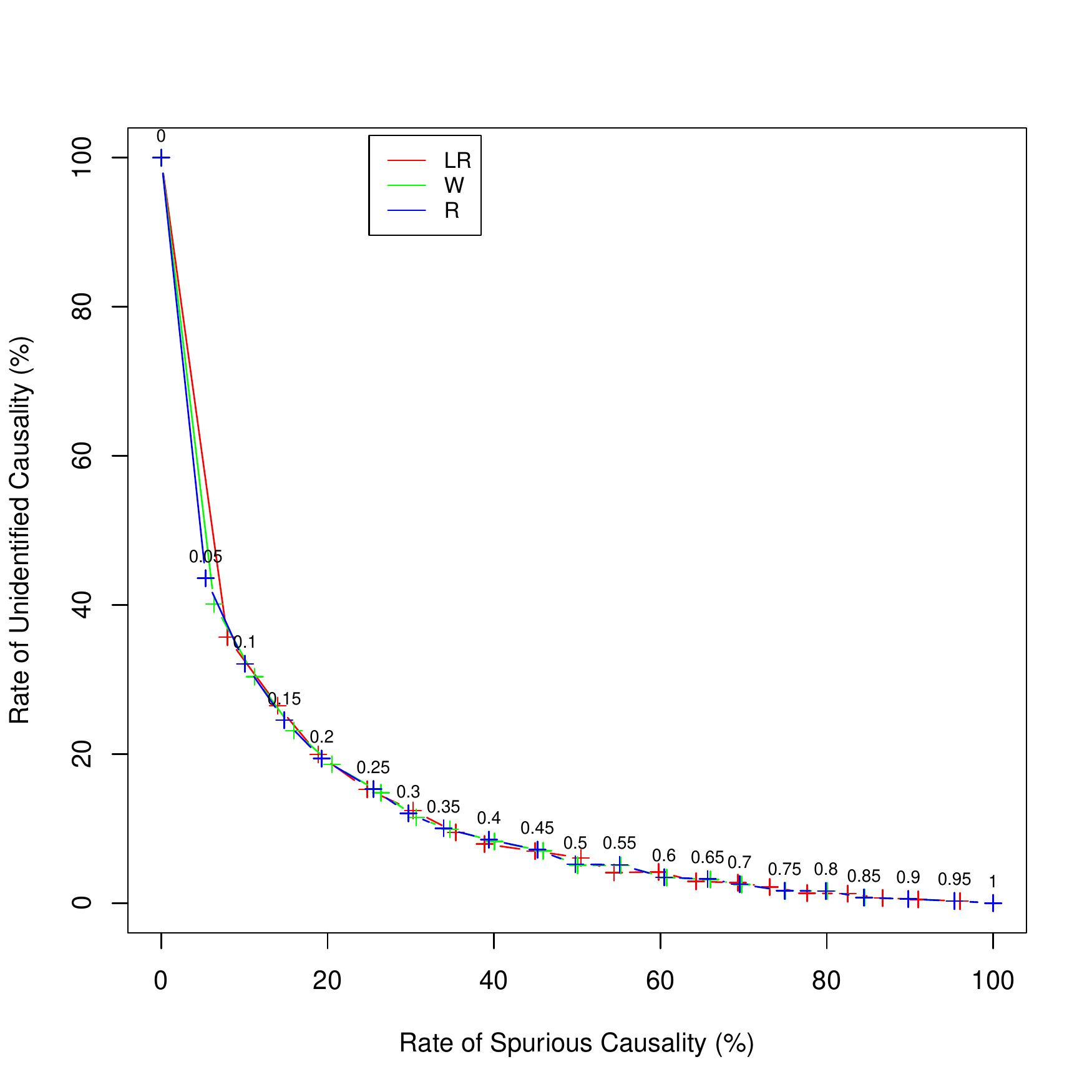}
 
   \vspace{-12pt}
    \caption{\it{\vspace{-6pt}Results for indirect topology with points closest to the ideal point of (0,0) corresponding to a significance level of 0.2.}} 
    \label{figa:unvsp_varA_indirect} 
\end{subfigure}

\vspace{-12pt}
\begin{subfigure}[b]{0.55\textwidth}
   \includegraphics[width=1\linewidth]{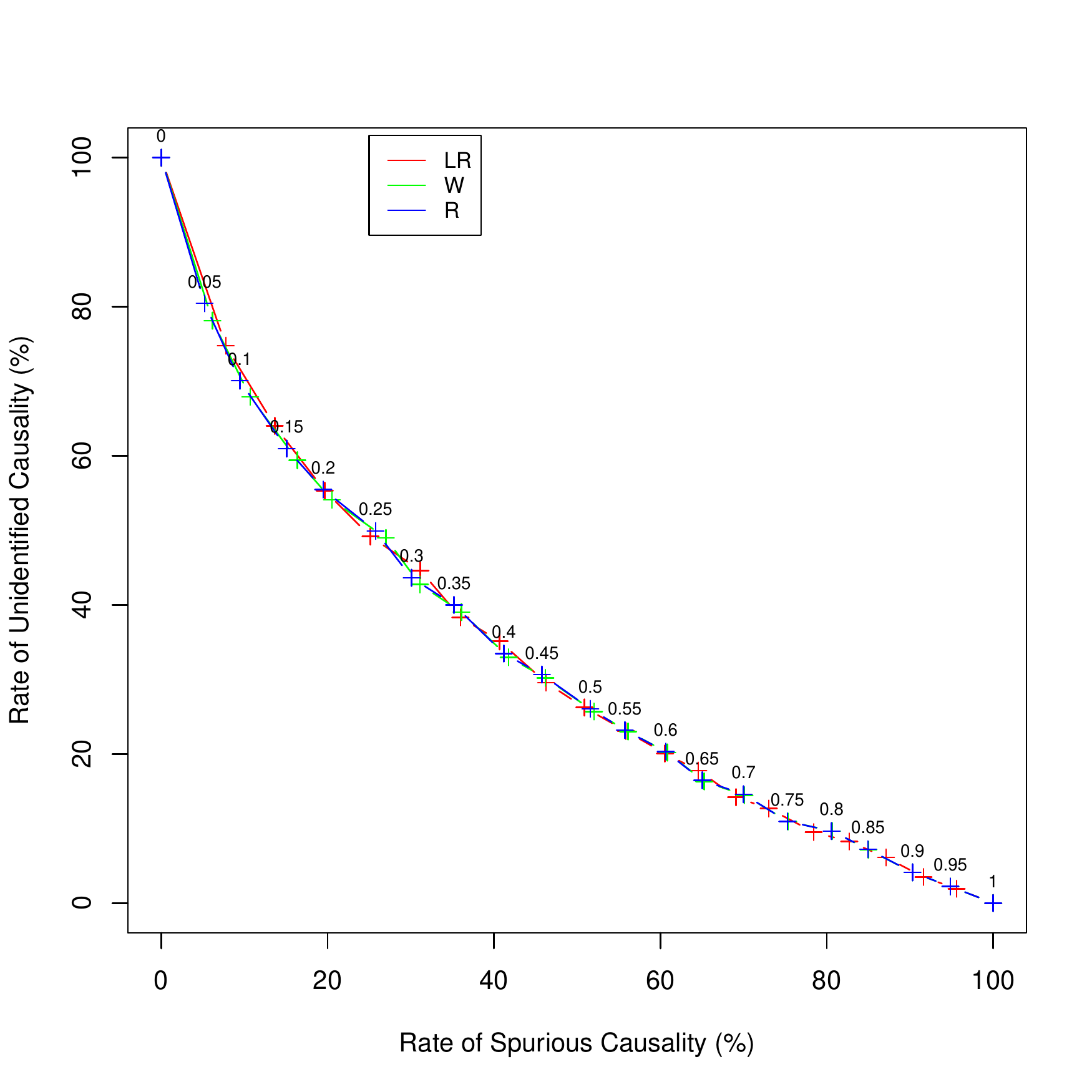}
   
   \vspace{-12pt}
    \caption{\it{\vspace{-6pt}Results for driver topology with points closest to the ideal point of (0,0) corresponding to a significance level of 0.3.}} 
    \label{figb:unvsp_varA_driver} 
\end{subfigure}
\caption{\it{Results for a Monte Carlo simulation using 5,000 iterations for indirect and driver topologies created by varying the significance level, for a fixed number of data points (50). Labels in the plot represent the significance level for the blue crosses (R test) with the following crosses for the green (W test) and red (LR test) correspond to the same significance level. The ideal point is represented in the bottom left corner as (0,0), where there is no unidentified or spurious causality.}}
\label{fig:unvsp_varA} 
\end{figure}

The first set of results shown here investigates varying the significance level (between $0$ and $1$) to determine the effect on the rate of unidentified causality and spurious causality for a low number of data points ($50$ data points was chosen following work by \cite{Taylor_1989}). The focus on a low number of data points is due to the deviation in prediction between LR, W and R, which (where we use a Chi-squared distribution with 2 degrees of freedom, and a look back of 2 time steps) is greatest at these lower values, hence optimising these results is of the most interest. The results shown in Fig.~\ref{fig:unvsp_varA} show that the best value for the significance level to optimise both the rate of unidentified causality and the rate of spurious causality are $0.2$ and $0.3$ for the indirect and driver topologies respectively. 

\begin{figure}
\centering

  \begin{subfigure}[b]{0.6\linewidth}
    \includegraphics[width=0.85\linewidth]{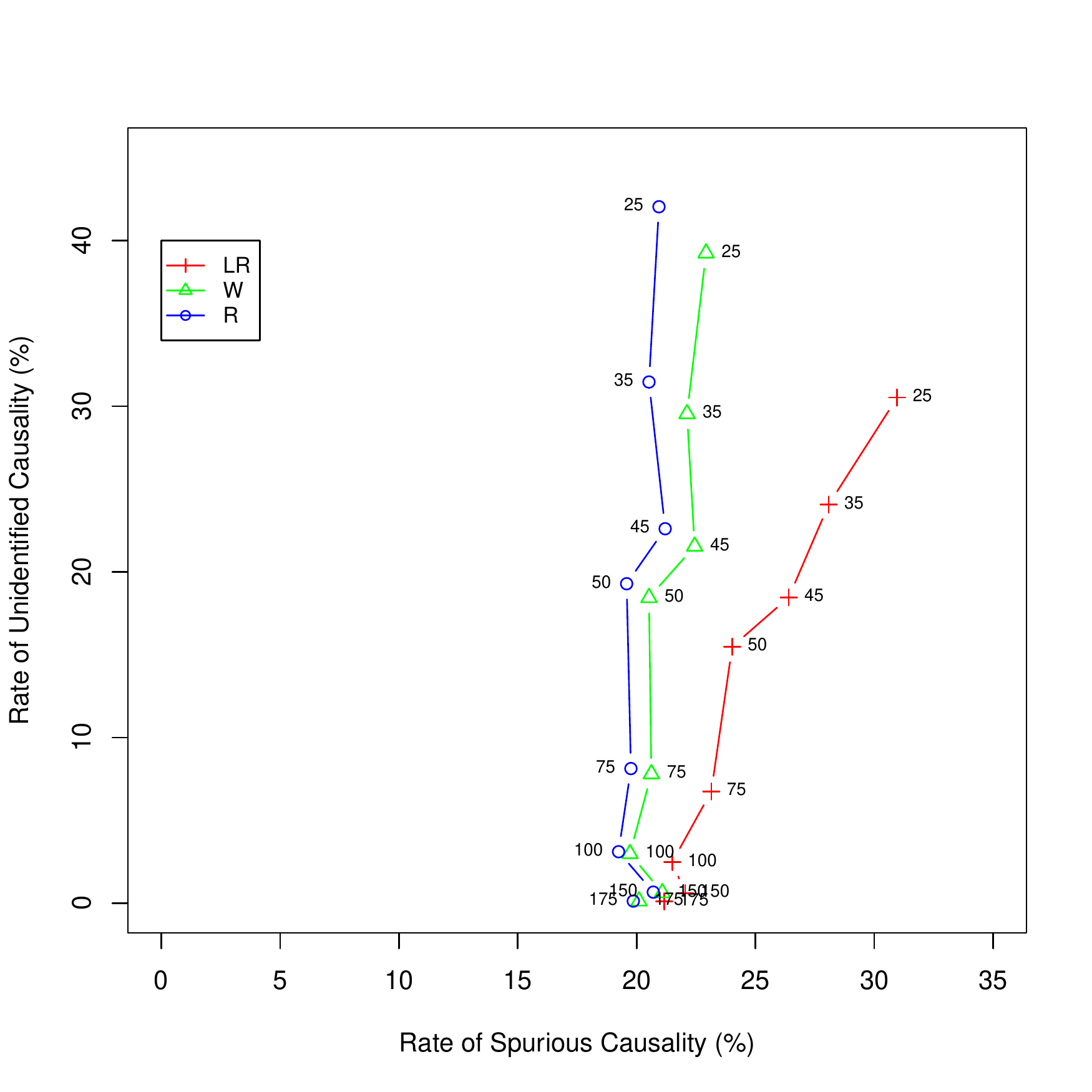} 
    
    \vspace{-10pt}
    \caption{\it{Indirect topology (significance level of 0.2). Maximum data points 175, where the rate of unidentified causality reaches zero.}} 
    \label{figa:unvssp_varN_indirect} 
  \end{subfigure}
  
  \vspace{-12pt}
  \begin{subfigure}[b]{0.6\linewidth}
    \includegraphics[width=0.85\linewidth]{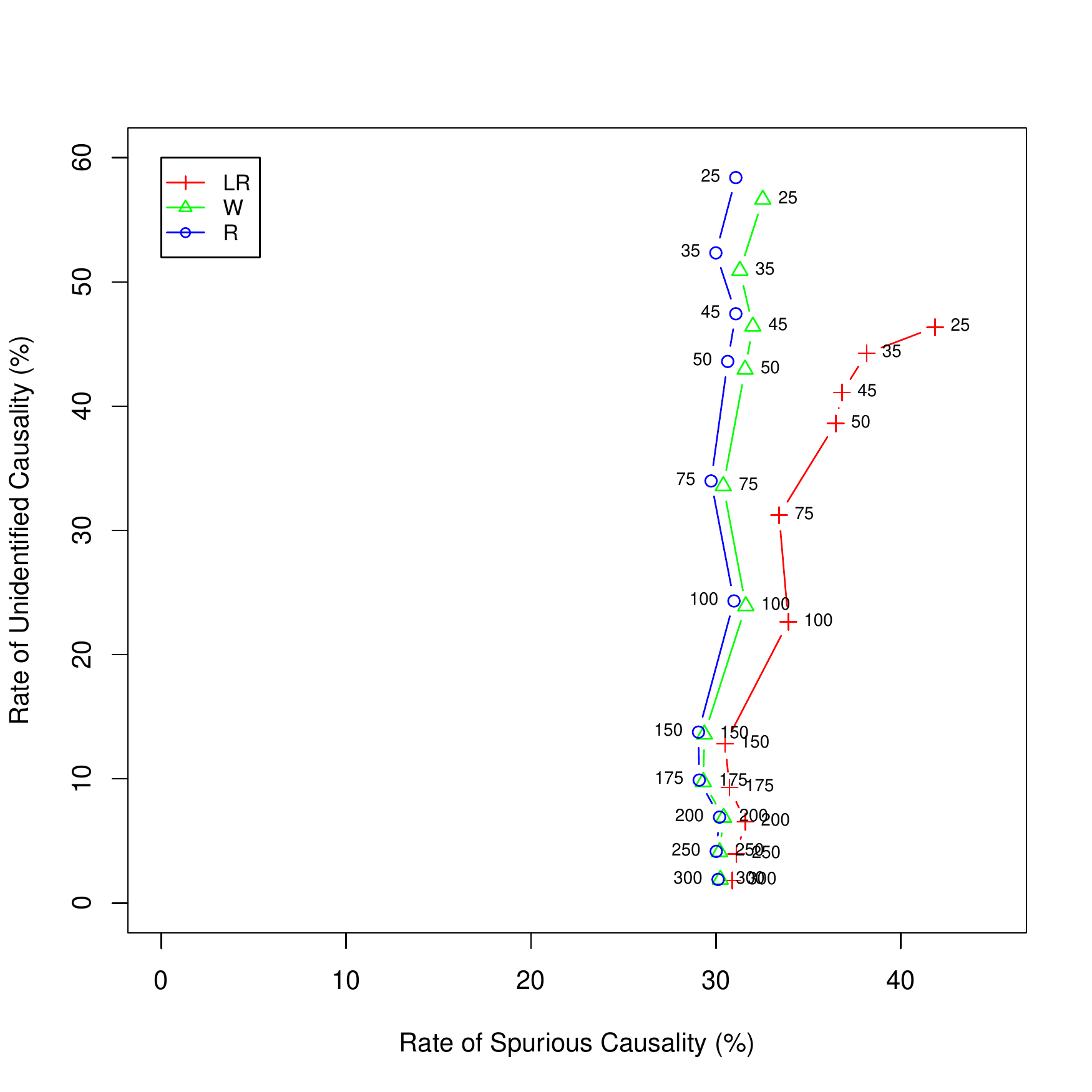} 
    
    \vspace{-10pt}
    \caption{\it{Driver topology (significance level of 0.3). Maximum data points 300, where the rate of unidentified causality reaches zero.}} 
    \label{figb:unvssp_varN_driver.pdf} 
  \end{subfigure}
     
  \caption{\it{Results for a Monte Carlo simulation using 5,000 iterations for an indirect and driver topology created by varying the number of data points, for a fixed significance level, 0.2 and 0.3 for an indirect and driver topology respectively. Labels in the plot represent the number of data points, with the number increasing down the plot. The ideal point is represented in the bottom left corner as (0,0), where there is no unidentified or spurious causality.}}
  \label{fig:unvssp_varN} 
\end{figure}

Using the significance levels which are closest to the most optimal point found for 50 data points in Figure~\ref{fig:unvsp_varA}, a Monte Carlo simulation was run with a varying number of data points, to investigate the behaviour of the three test criteria as the number of data points increased. For each sample size tested $1,000$ independently drawn test cases were run and average to give results. These results are shown in Figure~\ref{fig:unvssp_varN}, where the significance level is held constant ($0.2$ for the indirect topology and $0.3$ for the driver topology). The significance of the results presented in Figure~\ref{fig:unvssp_varN} was analysed using a hypothesis test and a significance level of $0.1$. Using this significance level it can be said that the  LR test produce results that are statically different to those produced by the W and R tests with regard to spurious causality for sample sizes up to and including $75$, and that it produces results for unidentified causality that are statically different for only sample sizes up to and including $25$ (with the exception of the driver topology where there is a statistical significance between the LR and R tests at a sample size of $35$). This hypothesis test also showed that there is no statistical difference between the W and R tests at any of the tested sample sizes. From these results it can be said that the three test criteria converge for the rate of spurious causality at a sample size of around $100$ and that before this convergence is reached the R and W tests offer the lowest rate of spurious causality, compared to the LR test. For very small sample sizes (i.e. around $25$) the LR test offers a lower rate of unidentified causality, compared to the R and W tests. These findings can be summarised as the W and R test being interchangeable and offer the results for spurious causality, and the LR test offering best results for unidentified causality at low sample size.

It should also be noted that the driver topology takes an increased number of data points to reach a zero rate of unidentified causality, demonstrating that the trivariate Granger analysis is more likely to have difficulty detecting a driver topology.

\clearpage

\section{Sensitivity of Trivariate Granger Analysis to Noise Terms} \label{sec:sensitivity_to_noise}

In the previous section, test cases were generated that contained a larger noise term on the $Z$ series than present on the $X$ or $Y$ series (using a standard deviation of $0.5$ opposed to $0.1$). Here we investigate how different combinations of the Signal-to-Noise ration (SNR) in the various time series affect the ability to infer causality in a trivariate Granger analysis.  Our initial investigation
considers the entire phase space, including contexts where a bivariate Granger analysis would not have inferred causality.

We define SNR in decibels \citep{Macmillan_1990} of the ratio of variances of signal and noise \citep{GrayNeuhoff_1998} as follows:
\begin{equation}
SNR = 10\log_{10} \left(  \frac{variance\; of\; the\; signal\; (noise\; free)}{variance\; of\; the\; noise} \right)
\end{equation}

For this investigation we will be investigating both intrinsic and extrinsic noise terms which we define as follows: 
\begin{itemize}
	\item {\it Intrinsic noise} - is the noise introduced within the functions used in the creation of the time series.
	\item {\it Extrinsic noise} - is the noise introduced when an external observer reads the data. This noise does not affect how the time series appear to other variables within the system ($X$, $Y$ and $Z$ will not see this noise), but an observer (e.g. who wishes to assess causality) will see the noise.
\end{itemize}

For a trivariate system the $SNR$ can be independently altered for each of the three present variables. Therefore comparing combinations of different $SNR$ values will produce a three-dimensional phase space, with each point corresponding to the either the rate of unidentified causality or the rate of spurious causality at that particular $SNR$ combination. For the following experiments we elected to use a range of $-40db$ to $40db$ for $SNR$ (this allowed for extremes to be investigated).  

\subsection{Intrinsic Noise}  \label{sec:intriniscNoise}   

For the experiment focussing on the intrinsic noise, we used a similar set up to the previous experiment (Section~\ref{sec:tc_testfunctions}), but with changes to the noise:

\begin{align*}
&x_{t} = U(-2, 2) + N(0, \alpha)\\
&y_{t} = 0.3\times y_{t-1} + x_{t-1} + N(0, \beta)\\
&z_{t} = 0.3\times z_{t-1} + x_{t-2} + N(0, \gamma)\\
&z_{t}^{'} = 0.3\times z_{t-1}^{'} + y_{t-1} + N(0, \gamma)\\
\end{align*}

Where $\alpha,\; \beta,$ and $\gamma$ are the standard deviation of the noise terms. A ``driver'' system uses $Z$ and an ``indirect'' system uses $Z^{'}$. The SNR of each time series will be given as $(SNR^{X},\; SNR^{Y},\; SNR^{Z})$
which are then converted into values for $(\alpha,\; \beta,\; \gamma)$.

Using the results from the previous section as guides we ran two sets of experiments, one at a high sample size ($300$ data points) and a low significance level ($0.05$), and one at a low sample size ($50$ data points) and a high significance level ($0.3$ for a driver topology, and $0.2$ for an indirect topology). In all of these experiments the W test criterion was used as it offers a lower rate of spurious causality than the LR test criterion (the R test criterion would also have been a valid choice). These two sets of tests were run with $500$ iterations, as an increase to $5,000$ iterations had very little effect on the main features of the graphs.

\begin{figure}
  \begin{subfigure}[b]{0.5\linewidth}
    \centering
    \includegraphics[width=1\linewidth]{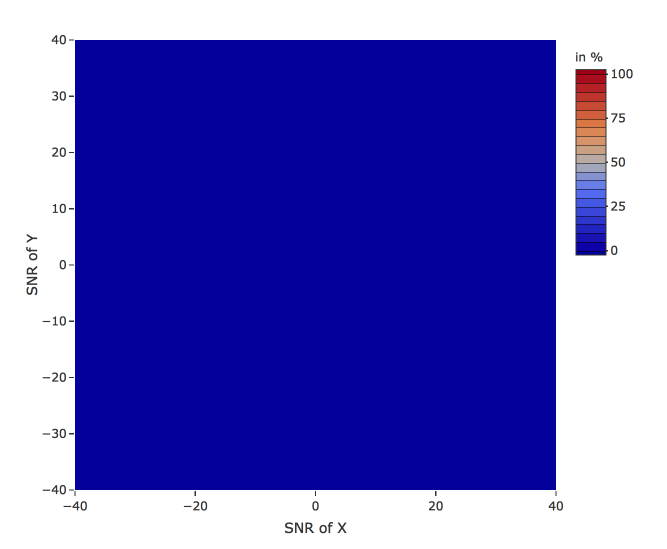} 
    \caption{\it{Rate of unidentified causality for $SNR^{Z} = 40db$.}} 
    \label{figa:topbot300d_untop} 
    \vspace{4ex}
  \end{subfigure}
  \begin{subfigure}[b]{0.5\linewidth}
    \centering
    \includegraphics[width=1\linewidth]{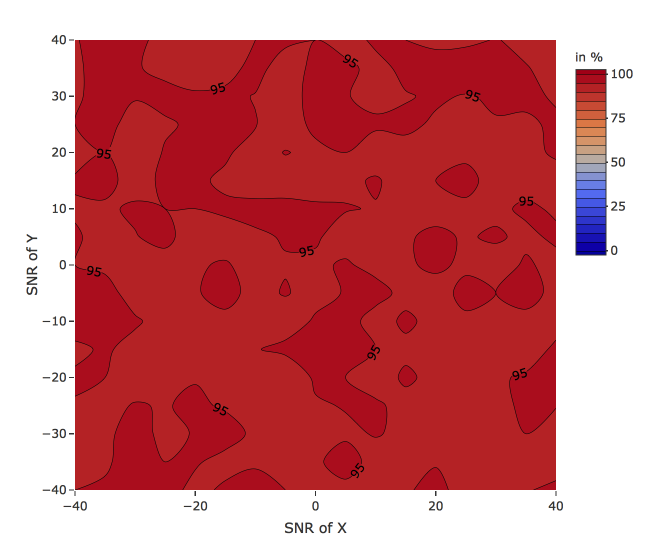} 
    \caption{\it{Rate of unidentified causality for $SNR^{Z} = -40db$.}} 
    \label{figb:topbot300d_unbot} 
    \vspace{4ex}
  \end{subfigure}
  
    \begin{subfigure}[b]{0.5\linewidth}
    \centering
    \includegraphics[width=1\linewidth]{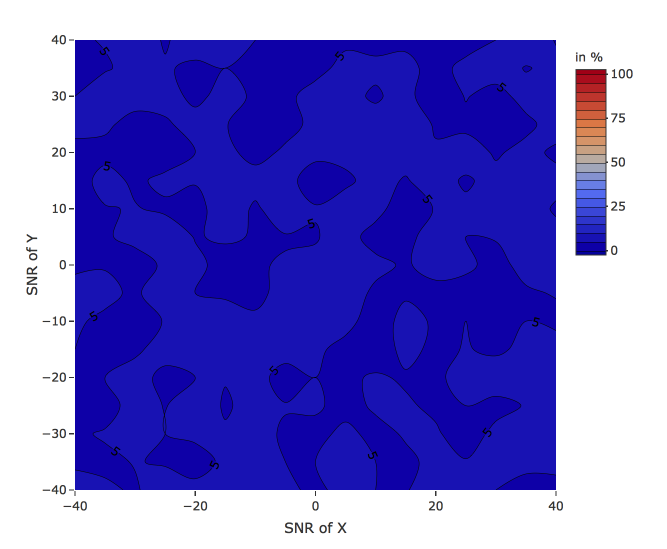} 
    \caption{\it{Rate of spurious causality for $SNR^{Z} = 40db$.}} 
    \label{figa:topbot300d_sptop} 
    \vspace{4ex}
  \end{subfigure}
  \begin{subfigure}[b]{0.5\linewidth}
    \centering
    \includegraphics[width=1\linewidth]{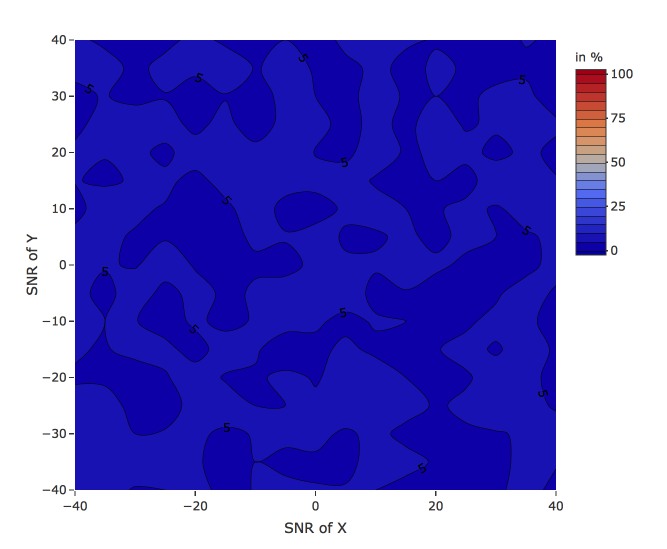} 
    \caption{\it{Rate of spurious causality for $SNR^{Z} = -40db$.}} 
    \label{figb:topbot300d_spbot} 
    \vspace{4ex}
  \end{subfigure}
  
  \caption{\it{Results for driver topology run with a sample of size of 300.}}
  \label{fig:topbot300d} 
\end{figure}

To understand the behaviour taking place within the phase space we chose to investigate a series of planes bisecting it, first two planes were selected for very low and very high $SNR^{Z}$ values ($-40db$ and $40db$ respectively), these results are shown in Figure~\ref{fig:topbot300d} for a sample size of $300$ and the driver case only, with the results for the indirect case being almost identical. 

\begin{figure}
  \begin{subfigure}[b]{0.5\linewidth}
    \centering
    \includegraphics[width=1\linewidth]{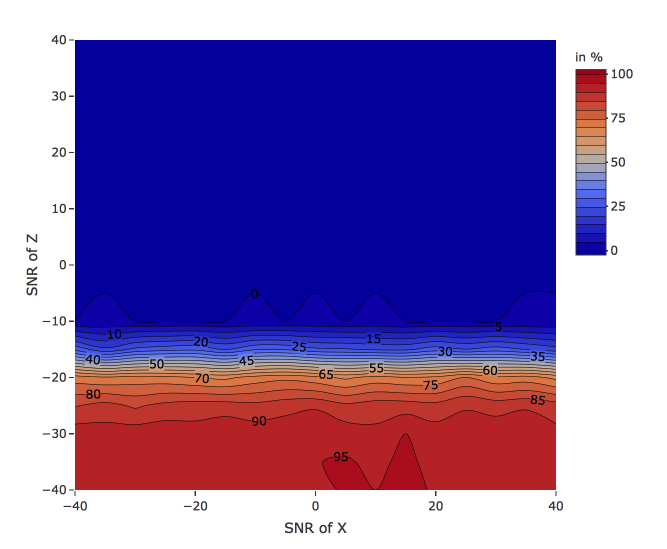} 
    \caption{\it{Rate of unidentified causality for\\ $SNR^{Y} = -40db$, in a driver topology.}} 
    \label{figa:front300D} 
    \vspace{4ex}
  \end{subfigure}
  \begin{subfigure}[b]{0.5\linewidth}
    \centering
    \includegraphics[width=1\linewidth]{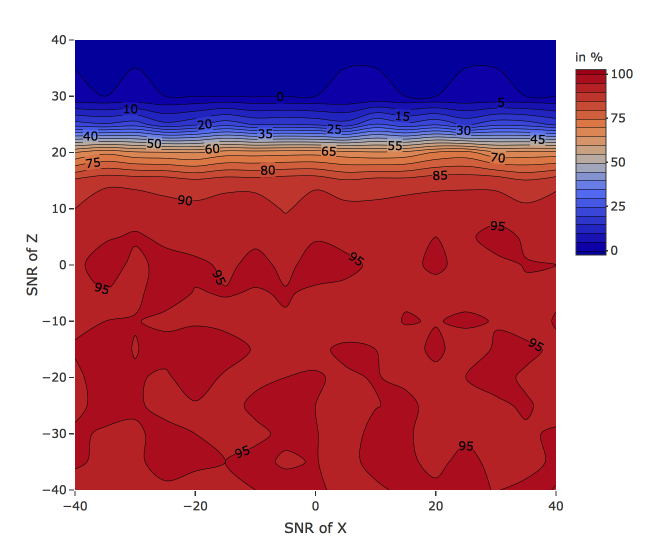} 
    \caption{\it{Rate of unidentified causality for\\ $SNR^{Y} = 40db$, in a driver topology.}} 
    \label{figb:back300D} 
    \vspace{4ex}
  \end{subfigure}
  
    \begin{subfigure}[b]{0.5\linewidth}
    \centering
    \includegraphics[width=1\linewidth]{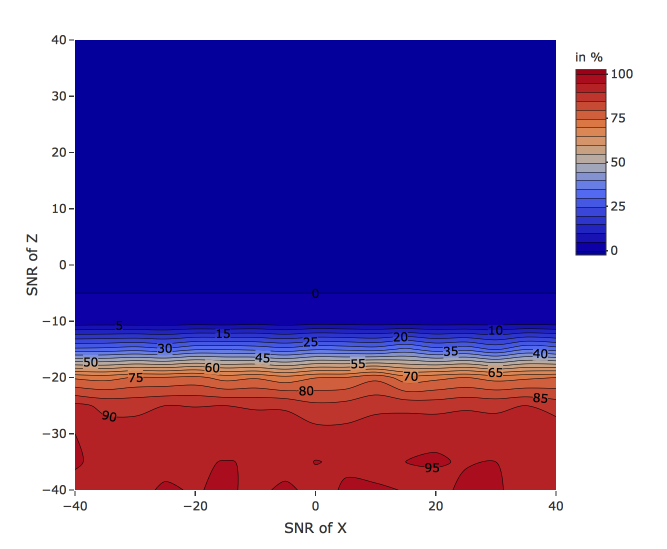} 
    \caption{\it{Rate of unidentified causality for\\ $SNR^{Y} = -40db$, in a indirect topology.}} 
    \label{figc:front300I} 
    \vspace{4ex}
  \end{subfigure}
  \begin{subfigure}[b]{0.5\linewidth}
    \centering
    \includegraphics[width=1\linewidth]{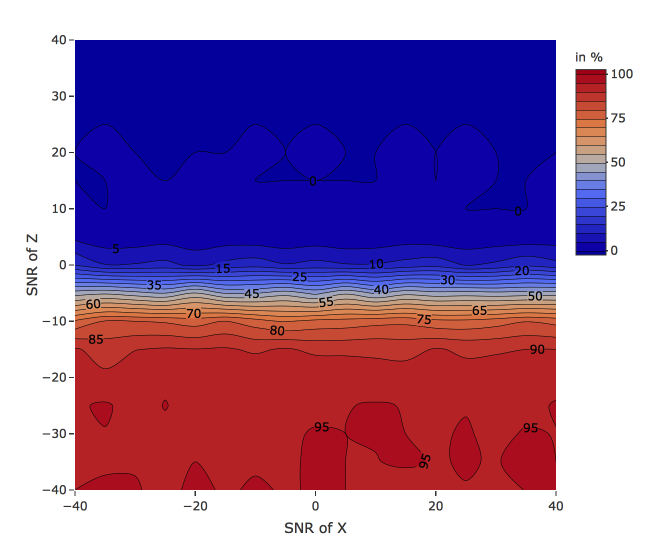} 
    \caption{\it{Rate of unidentified causality for\\ $SNR^{Y} = 40db$, in a indirect topology.}} 
    \label{figd:back300I} 
    \vspace{4ex}
  \end{subfigure}
  
  \caption{\it{Results for driver and indirect topology run with a sample of size of 300.}}
  \label{fig:frontback300id} 
\end{figure}

\begin{figure}
  \begin{subfigure}[b]{0.5\linewidth}
    \centering
    \includegraphics[width=1\linewidth]{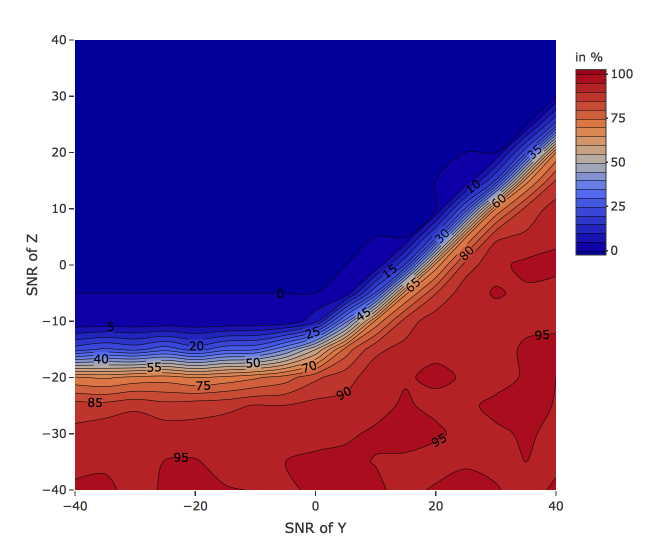} 
    \caption{\it{Rate of unidentified causality for\\ $SNR^{X} = -40db$, in a driver topology.}} 
    \label{figa:left300D} 
    \vspace{4ex}
  \end{subfigure}
  \begin{subfigure}[b]{0.5\linewidth}
    \centering
    \includegraphics[width=1\linewidth]{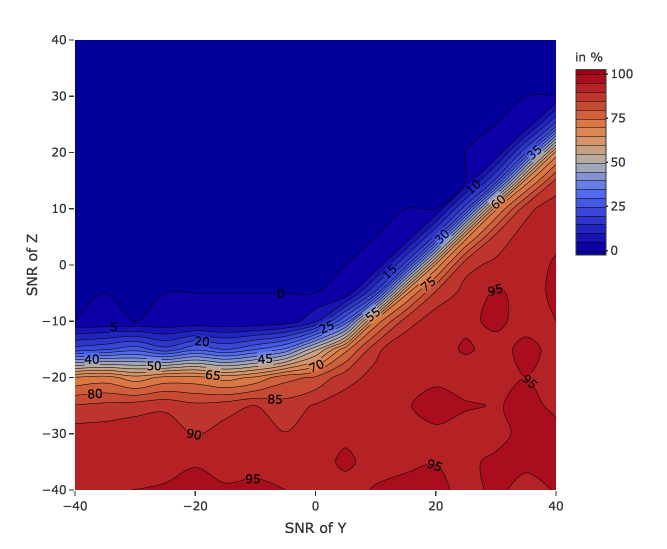} 
    \caption{\it{Rate of unidentified causality for\\ $SNR^{X} = 40db$, in a driver topology.}} 
    \label{figb:right300D} 
    \vspace{4ex}
  \end{subfigure}
  
    \begin{subfigure}[b]{0.5\linewidth}
    \centering
    \includegraphics[width=1\linewidth]{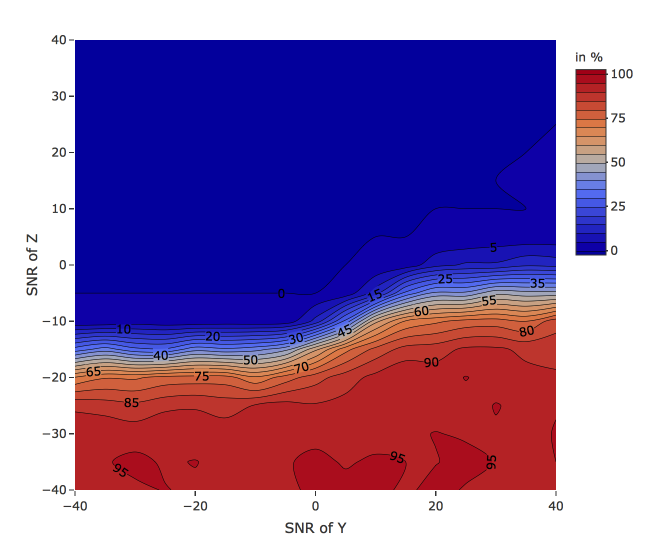} 
    \caption{\it{Rate of unidentified causality for\\ $SNR^{X} = -40db$, in a indirect topology.}} 
    \label{figc:left300I} 
    \vspace{4ex}
  \end{subfigure}
  \begin{subfigure}[b]{0.5\linewidth}
    \centering
    \includegraphics[width=1\linewidth]{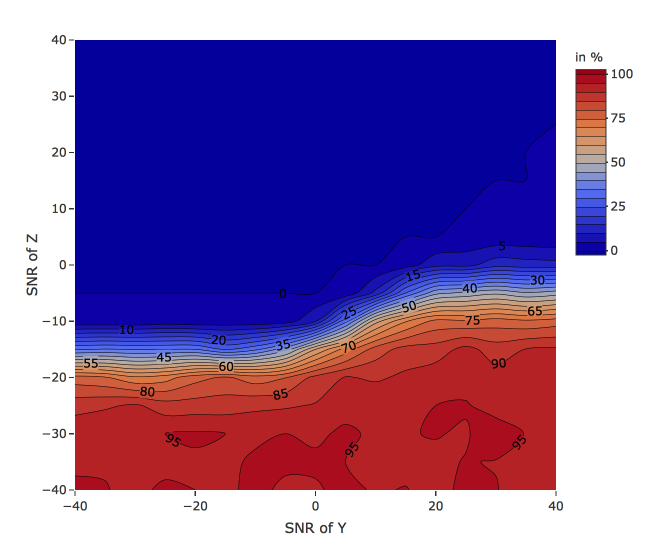} 
    \caption{\it{Rate of unidentified causality for\\ $SNR^{X} = 40db$, in a indirect topology.}} 
    \label{figd:right300I} 
    \vspace{4ex}
  \end{subfigure}
  
  \caption{\it{Results for driver and indirect topology run with a sample of size of 300.}}
  \label{fig:leftright300id} 
\end{figure}

Figures~\ref{figa:topbot300d_untop}~and~\ref{figb:topbot300d_unbot} demonstrate a strong polarisation between the rate of unidentified causality for a low value and high value of $SNR^{Z}$. This polarisation occurs due to a shift in the behaviour of the system along the $Z$ axis, moving from a zero rate (Figure~\ref{figa:topbot300d_untop}) to a high rate (Figure~\ref{figb:topbot300d_unbot}) of unidentified causality. To characterise this shift four planes that follow the $Z$ axis were plotted, these were for both a low ($-40db$) and high ($40db$) fixed $SNR^{Y}$ and a low ($-40db$) and high ($40db$) fixed $SNR^{X}$, shown in Figures~\ref{fig:frontback300id}~and~\ref{fig:leftright300id}  respectively. 

From Figures~\ref{figa:topbot300d_sptop} and \ref{figb:topbot300d_spbot} it can be seen that for the two cases presented the rate of spurious causality remains approximately uniform and tends towards the expected theoretical value, $5\%$. These results, along with investigation of other planes in the phase space (not shown here), allow us to conclude that the rate of spurious causality is largely unaffected by the value of the $SNR$ on any of the present variables.

Combing the results from Figures~\ref{fig:frontback300id} and \ref{fig:leftright300id} one can see the outline of a gradual transition from relatively low rates of unidentified causality to relatively high rates of unidentified causality that takes place through the $Z$ axis. Figure~\ref{fig:frontback300id} shows that the value of $SNR^{X}$ has little impact on this transition, with Figure~\ref{fig:leftright300id} demonstrating that for negative values of $SNR^{Y}$, $SNR^{Z}$ determines the transition. As $SNR^{Y}$ becomes positive (as illustrated in Figure~\ref{fig:leftright300id}) it alters the effect of $SNR^{Z}$ on the transition, causing it to occur at higher levels of $SNR^{Z}$. Though this behaviour is present in both the driver and indirect topology, it is much more pronounced in the driver case, therefore the overall likelihood of having unidentified causality is higher in the driver topology then in the indirect topology (assuming that all points in the phase space are equally likely). 

\begin{figure}
  \begin{subfigure}[b]{0.5\linewidth}
    \centering
    \includegraphics[width=1\linewidth]{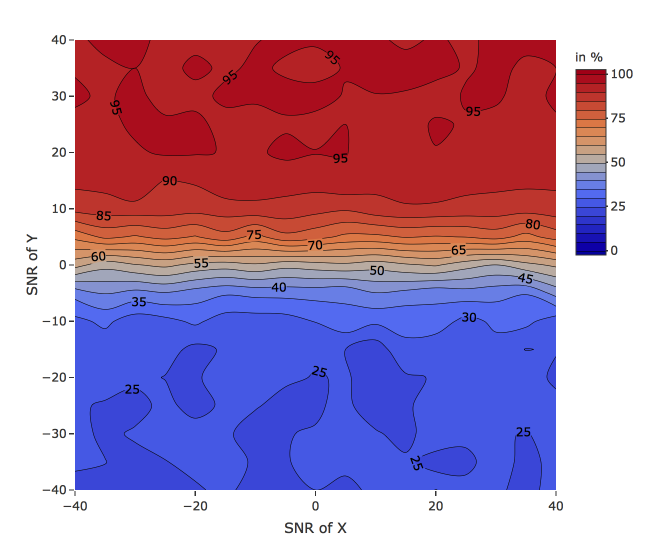} 
    \caption{\it{Rate of unidentified causality for\\ $SNR^{Z} = -15db$, in a driver topology.}} 
    \label{figa:left300D} 
    \vspace{4ex}
  \end{subfigure}
  \begin{subfigure}[b]{0.5\linewidth}
    \centering
    \includegraphics[width=1\linewidth]{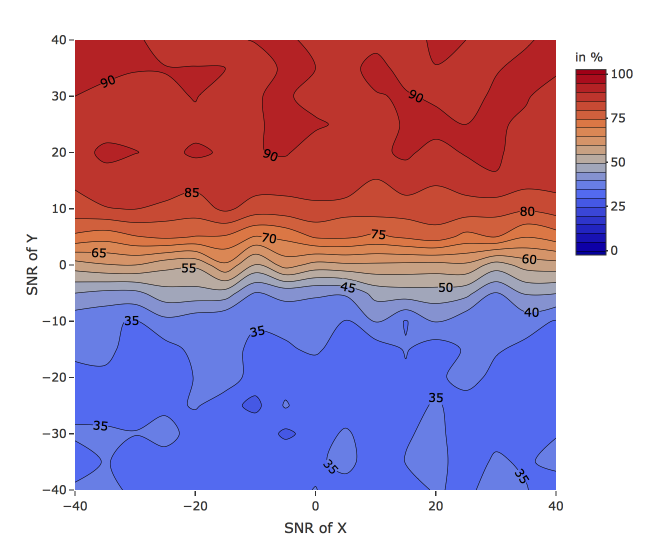} 
    \caption{\it{Rate of unidentified causality for\\ $SNR^{Z} = -15db$, in a indirect topology.}} 
    \label{figb:right300D} 
    \vspace{4ex}
  \end{subfigure}

  \caption{\it{Results for driver and indirect topology run with a sample of size of 300.}}
  \label{fig:mid300id} 
\end{figure}

So far this transition has been characterised through a perimeter defined at the extreme values ($-40db$ and $40db$), to further define this characterisation and to investigate the surface of this shift, a plane at $SNR^{Z} = -15$ is shown in Figure~\ref{fig:mid300id}.   

Figure~\ref{fig:mid300id} demonstrates that this transition takes place approximately smoothly across the $SNR^{X}$ axis and hence there is no behaviour present that would not be expected from investigation of its extreme values (Figures~\ref{fig:frontback300id} and \ref{fig:leftright300id}). 

\begin{figure}
  \begin{subfigure}[b]{0.5\linewidth}
    \centering
    \includegraphics[width=1\linewidth]{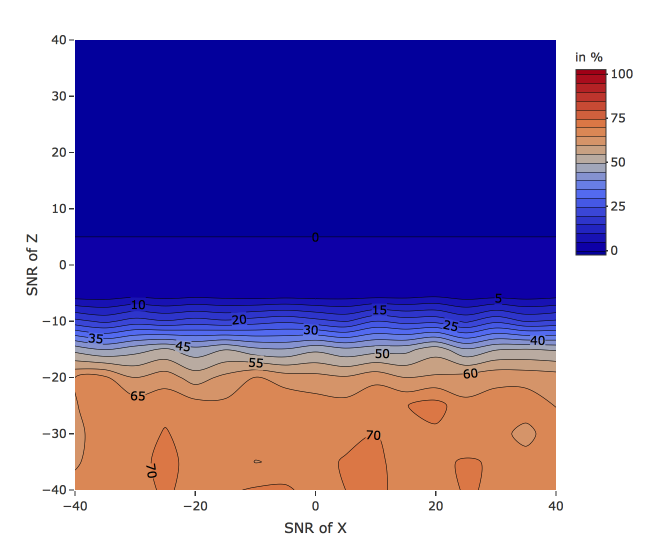} 
    \caption{\it{Rate of unidentified causality for\\ $SNR^{Y} = -40db$, in a driver topology.}} 
    \label{figa:front50D} 
    \vspace{4ex}
  \end{subfigure}
  \begin{subfigure}[b]{0.5\linewidth}
    \centering
    \includegraphics[width=1\linewidth]{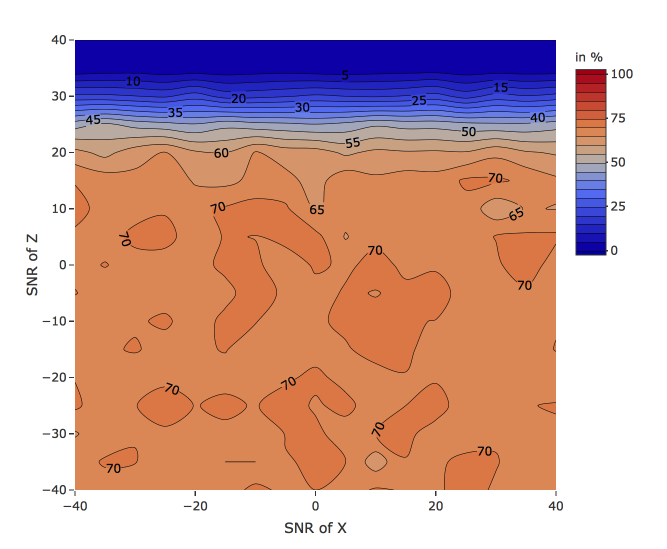} 
    \caption{\it{Rate of unidentified causality for\\ $SNR^{Y} = 40db$, in a driver topology.}} 
    \label{figb:back50D} 
    \vspace{4ex}
  \end{subfigure}
  
    \begin{subfigure}[b]{0.5\linewidth}
    \centering
    \includegraphics[width=1\linewidth]{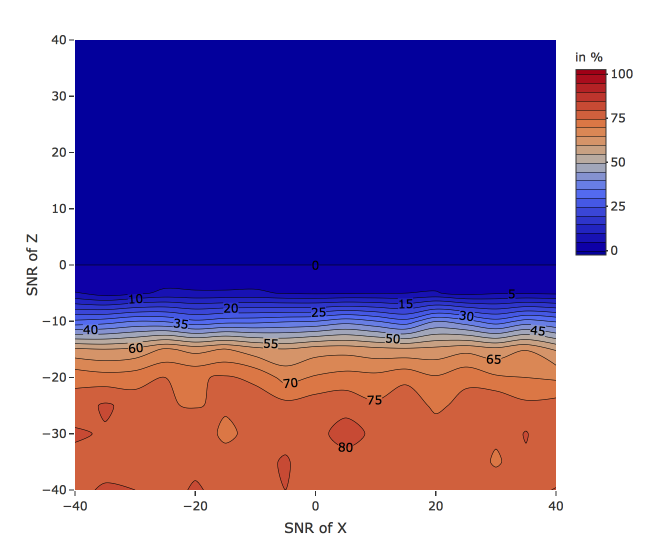} 
    \caption{\it{Rate of unidentified causality for\\ $SNR^{Y} = -40db$, in a indirect topology.}} 
    \label{figc:front50I} 
    \vspace{4ex}
  \end{subfigure}
  \begin{subfigure}[b]{0.5\linewidth}
    \centering
    \includegraphics[width=1\linewidth]{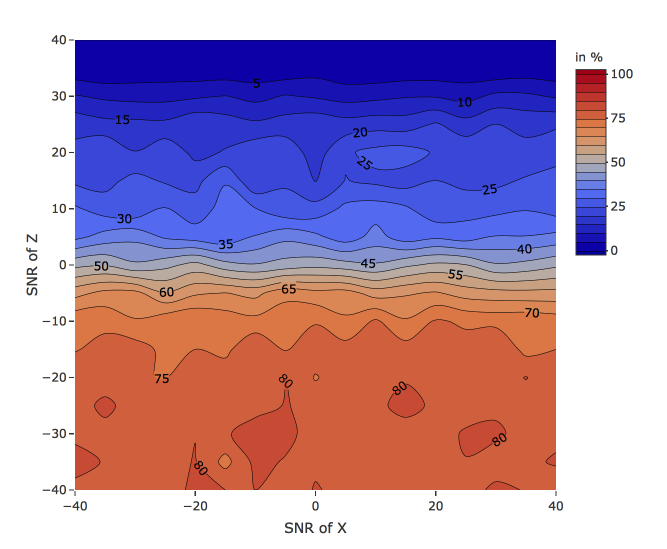} 
    \caption{\it{Rate of unidentified causality for\\ $SNR^{Y} = 40db$, in a indirect topology.}} 
    \label{figd:back50I} 
    \vspace{4ex}
  \end{subfigure}
  
  \caption{\it{Results for driver and indirect topology run with a sample of size of 50.}}
  \label{fig:frontback50id} 
\end{figure}

\begin{figure}
  \begin{subfigure}[b]{0.5\linewidth}
    \centering
    \includegraphics[width=1\linewidth]{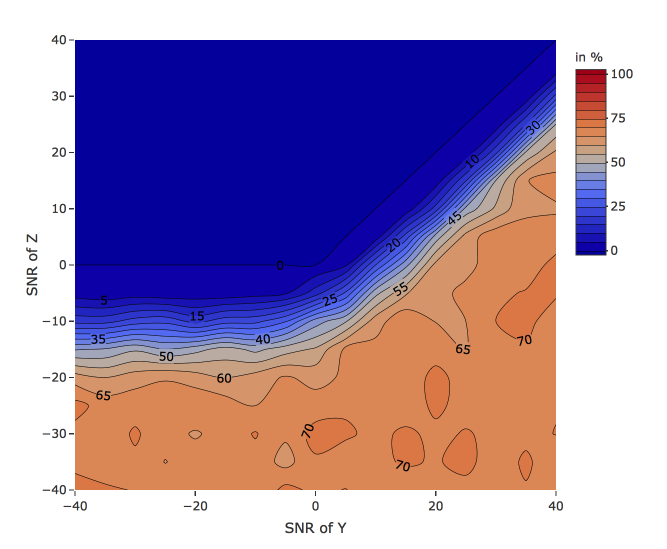} 
    \caption{\it{Rate of unidentified causality for\\ $SNR^{X} = -40db$, in a driver topology.}} 
    \label{figa:left50D} 
    \vspace{4ex}
  \end{subfigure}
  \begin{subfigure}[b]{0.5\linewidth}
    \centering
    \includegraphics[width=1\linewidth]{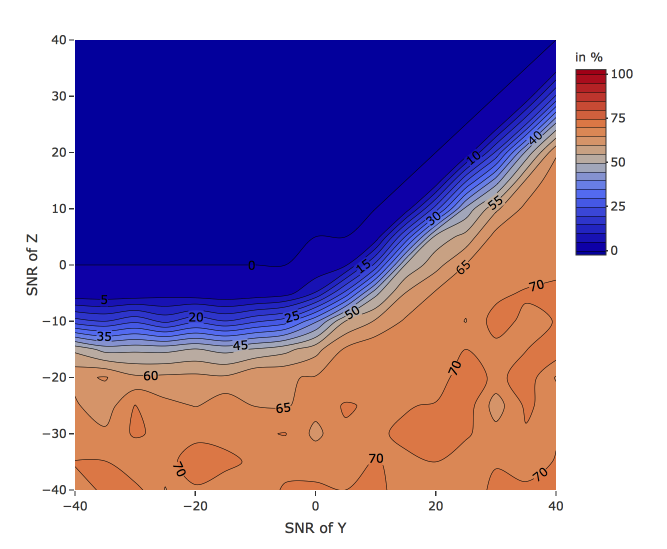} 
    \caption{\it{Rate of unidentified causality for\\ $SNR^{X} = 40db$, in a driver topology.}} 
    \label{figb:right50D} 
    \vspace{4ex}
  \end{subfigure}
  
    \begin{subfigure}[b]{0.5\linewidth}
    \centering
    \includegraphics[width=1\linewidth]{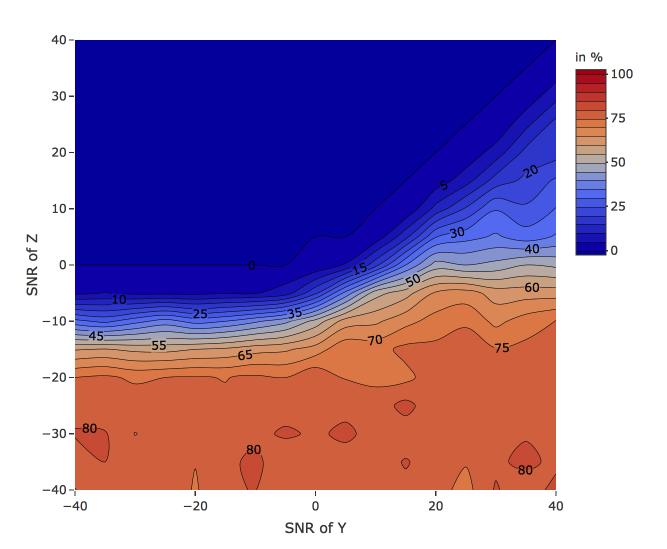} 
    \caption{\it{Rate of unidentified causality for\\ $SNR^{X} = -40db$, in a indirect topology.}} 
    \label{figc:left50I} 
    \vspace{4ex}
  \end{subfigure}
  \begin{subfigure}[b]{0.5\linewidth}
    \centering
    \includegraphics[width=1\linewidth]{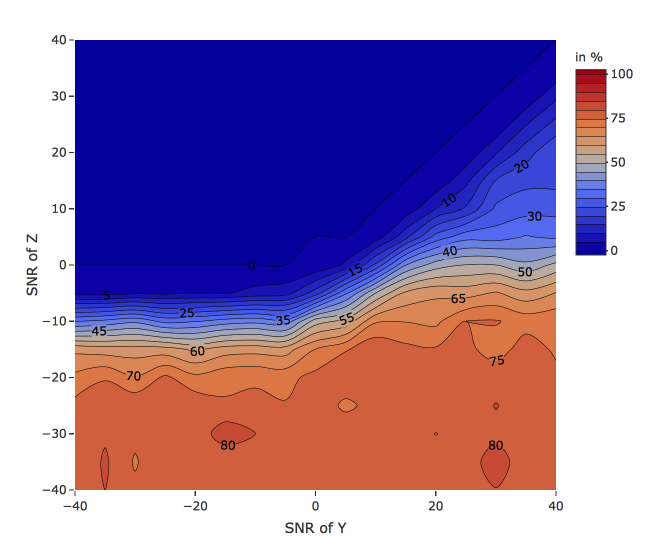} 
    \caption{\it{Rate of unidentified causality for\\ $SNR^{X} = 40db$, in a indirect topology.}} 
    \label{figd:right50I} 
    \vspace{4ex}
  \end{subfigure}
  
  \caption{\it{Results for driver and indirect topology run with a sample of size of 50.}}
  \label{fig:leftright50id} 
\end{figure}

The behaviour of the phase space for a smaller sample size ($50$) is similar, with the rate of spurious causality being constant around the theoretical expected value ($30\%$ and $20\%$ for a driver and indirect topology respectively). The perimeter of the transition within the unidentified causality phase space can be seen in Figures~\ref{fig:frontback50id} and \ref{fig:leftright50id}, where it can be observed that a zero rate of unidentified causality is reached at a higher value of $SNR^{Z}$ (e.g compares Figures \ref{figa:front300D} and \ref{figa:front50D}), but that the comparative rates are lower than in the $300$ sample size case (this is due to the higher significance level). Though not shown the transition takes place relatively smoothly across the $SNR^{X}$ axis similar to the results shown in Figure~\ref{fig:mid300id}.

\clearpage

\subsection{Extrinsic Noise} \label{sec:extrinsiNoise}

The above experiments using intrinsic noise were then repeated for a system containing extrinsic noise. For this set-up the equations that were used for creation of the time series were the following:

\vspace{-24pt}
\begin{align*}
&x_{t}^{'} = U(-2, 2)\\
&y_{t}^{'} = 0.3\times y_{t-1}^{'} + x_{t-1}\\
&z_{t}^{'} = 0.3\times z_{t-1}^{'} + x_{t-2}\\
&z_{t}^{''} = 0.3\times z_{t-1}^{''} + y_{t-1}\\
\end{align*}

\vspace{-24pt}
These equations follow the same notation as those above, but now the extrinsic noise is added post creation of the time series, as follows.

\vspace{-24pt}
\begin{align*}
&x_{t} = x_{t}^{'} + N(0, \alpha)\\
&y_{t} = y_{t}^{'} + N(0, \beta)\\
&z_{t} = z_{t}^{'} + N(0, \gamma)\\
&z_{t}^{'} = z_{t}^{''} + N(0, \gamma)\\
\end{align*}

\vspace{-24pt}
The definitions of $\alpha,\; \beta,$ and $\gamma$ and of $SNR^{X},\; SNR^{Y},$ and $SNR^{Z}$, are the same as for the previous experimental set-up, with each run again being conducted using an iteration size of $500$. 

The phase space observations rely on an understanding of the actual underlying topology (which would not normally be known).  For the two underlying topologies (indirect and driver) the trivariate Granger results are assessed to determine the extent to which they exhibit either spurious causality or unidentified causality.  Each of these four cases is effectively observing the probability that a specific link is determined to be causal; Table~\ref{tbl:extrinsic_observations} lists the key link for each case and states which figure illustrates the relevant phase space observations (planes that contain very little variation are not presented).

\begin{table}[h]
\centering
\caption{Key links for phase space observations.\label{tbl:extrinsic_observations}}
\begin{tabular}{ l c c }
       &Spurious causality & Unidentified causality \\
Driver topology& $Y \rightarrow Z$ (Figure~\ref{fig:all300spD})& $X \rightarrow Z$ (Figure~\ref{fig:allt300unD}) \\
Indirect topology& $X \rightarrow Z$ (Figure~\ref{fig:all300spI})& $Y \rightarrow Z$ (Figure~\ref{fig:all300unI}) \\
\end{tabular}
\end{table}

\begin{figure}
  \begin{subfigure}[b]{0.5\linewidth}
    \centering
    \includegraphics[width=1\linewidth]{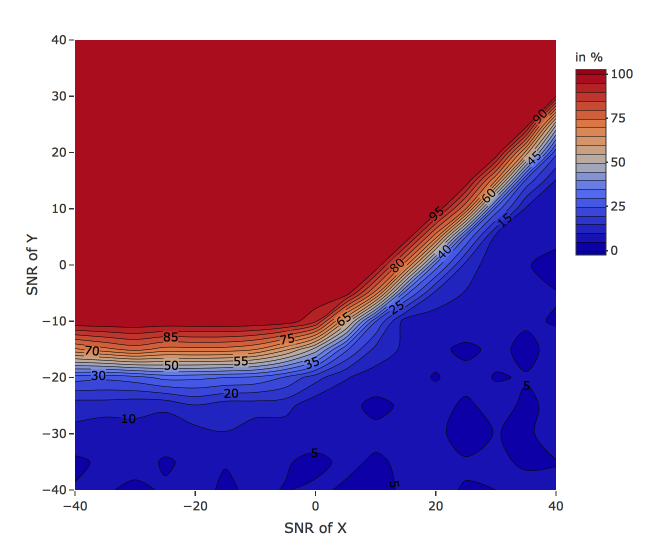} 
    \caption{\it{Rate of spurious causality for\\ $SNR^{Z} = 40db$, in a driver topology.}} 
    \label{figa:Ztop_sp300D} 
    \vspace{4ex}
  \end{subfigure}
  \begin{subfigure}[b]{0.5\linewidth}
    \centering
    \includegraphics[width=1\linewidth]{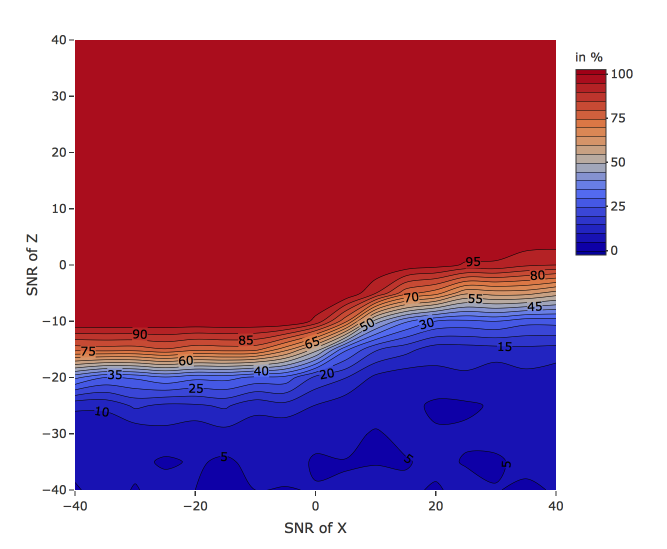} 
    \caption{\it{Rate of spurious causality for\\ $SNR^{Y} = 40db$, in a driver topology.}} 
    \label{figb:Yback_sp300D} 
    \vspace{4ex}
  \end{subfigure}
  
    \begin{subfigure}[b]{0.5\linewidth}
    \centering
    \includegraphics[width=1\linewidth]{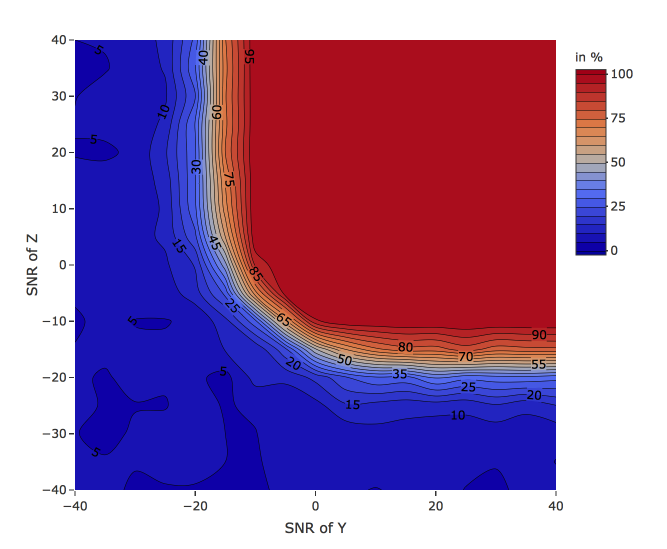} 
    \caption{\it{Rate of spurious causality for\\ $SNR^{X} = -40db$, in a driver topology.}} 
    \label{figc:Xleft_sp300D} 
    \vspace{4ex}
  \end{subfigure}
  \begin{subfigure}[b]{0.5\linewidth}
    \centering
    \includegraphics[width=1\linewidth]{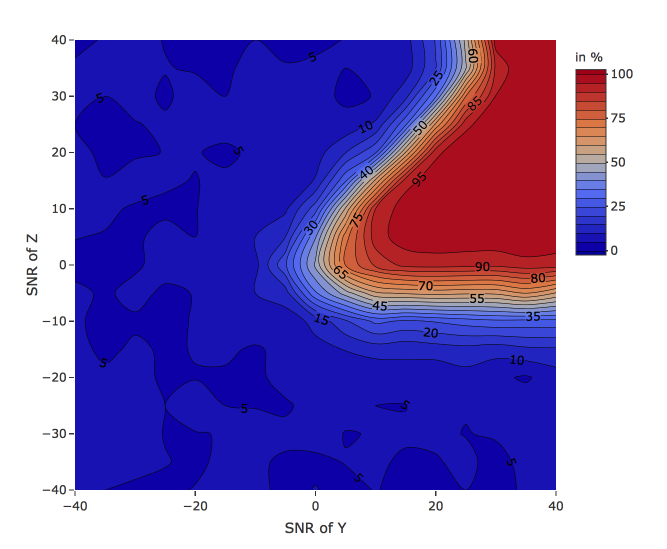} 
    \caption{\it{Rate of spurious causality for\\ $SNR^{X} = 40db$, in a driver topology.}} 
    \label{figd:Xright_sp300D} 
    \vspace{4ex}
  \end{subfigure}
  
  \caption{\it{Results for driver topology run with a sample of size of 300, encompassing the transition from from a low to high rate of spurious causality.}}
  \label{fig:all300spD} 
\end{figure}

\begin{figure}
  \begin{subfigure}[b]{0.5\linewidth}
    \centering
    \includegraphics[width=1\linewidth]{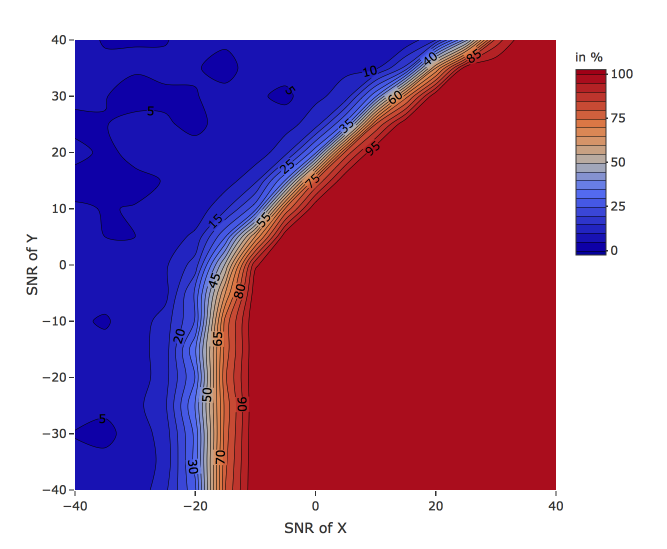} 
    \caption{\it{Rate of spurious causality for\\ $SNR^{Z} = 40db$, in a indirect topology.}} 
    \label{figa:Ztop_sp300I} 
    \vspace{4ex}
  \end{subfigure}
  \begin{subfigure}[b]{0.5\linewidth}
    \centering
    \includegraphics[width=1\linewidth]{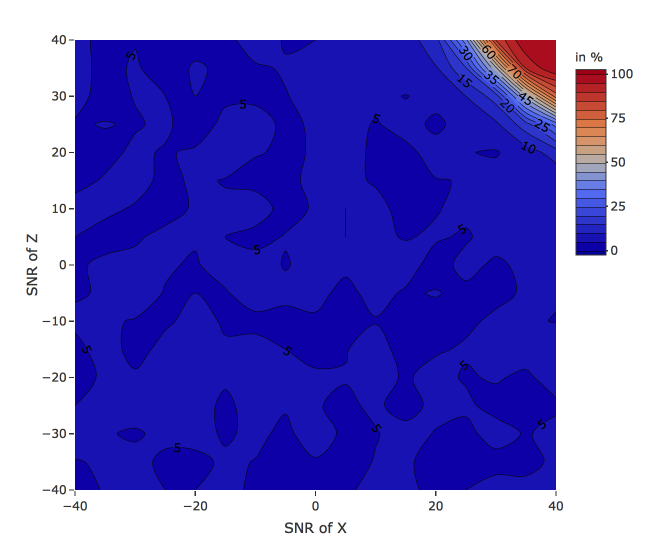} 
    \caption{\it{Rate of spurious causality for\\ $SNR^{Y} = 40db$, in a indirect topology.}} 
    \label{figb:Yback_sp300I} 
    \vspace{4ex}
  \end{subfigure}
  
    \begin{subfigure}[b]{0.5\linewidth}
    \centering
    \includegraphics[width=1\linewidth]{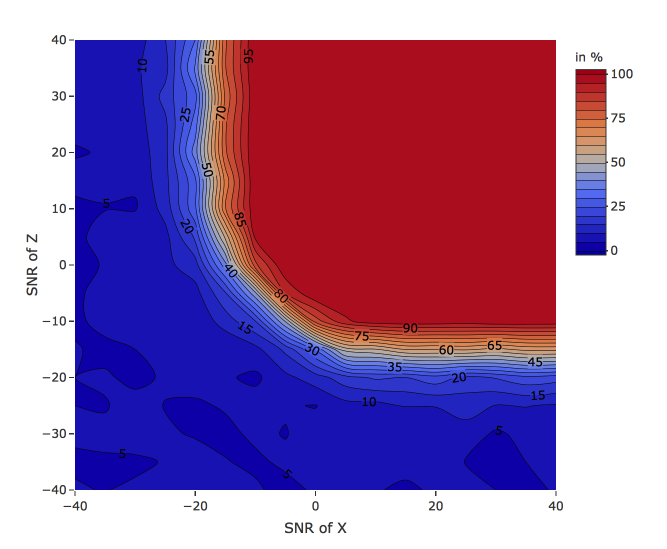} 
    \caption{\it{Rate of spurious causality for\\ $SNR^{Y} = -40db$, in a indirect topology.}} 
    \label{figc:Yfront_sp300I} 
    \vspace{4ex}
  \end{subfigure}
  \begin{subfigure}[b]{0.5\linewidth}
    \centering
    \includegraphics[width=1\linewidth]{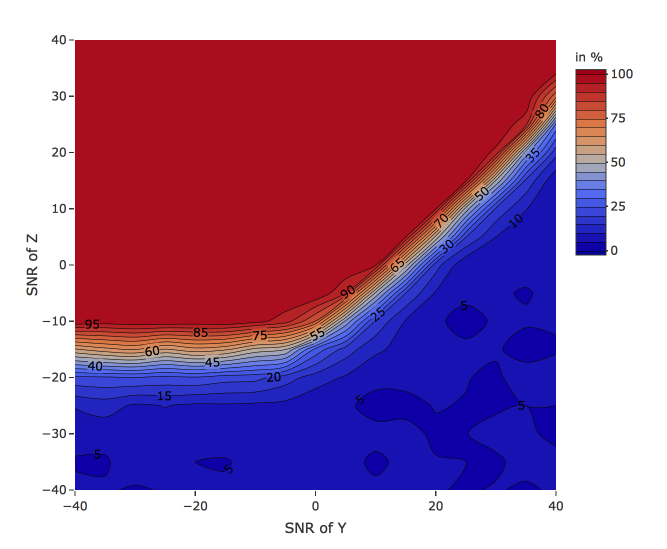} 
    \caption{\it{Rate of spurious causality for\\ $SNR^{X} = 40db$, in a indirect topology.}} 
    \label{figd:Xright_sp300I} 
    \vspace{4ex}
  \end{subfigure}
  
  \caption{\it{Results for indirect topology run with a sample of size of 300, encompassing the transition from from a low to high rate of spurious causality.}}
  \label{fig:all300spI} 
\end{figure}

\begin{figure}
  \begin{subfigure}[b]{0.5\linewidth}
    \centering
    \includegraphics[width=1\linewidth]{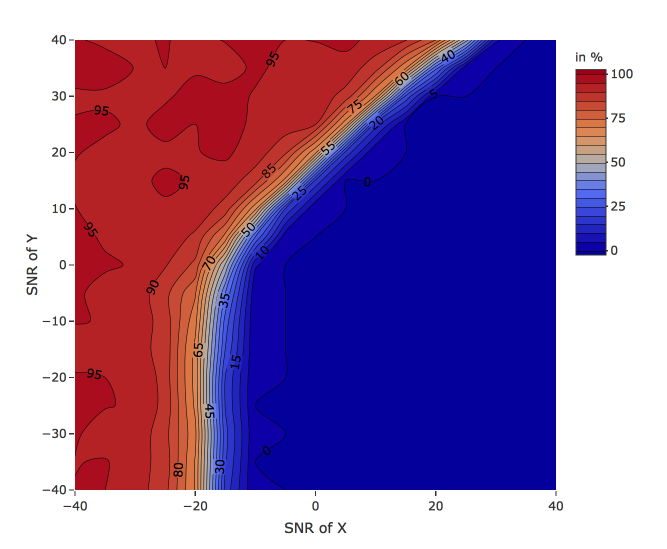} 
    \caption{\it{Rate of unidentified causality for\\ $SNR^{Z} = 40db$, in a driver topology.}} 
    \label{figa:Ztop_un300D} 
    \vspace{4ex}
  \end{subfigure}
  \begin{subfigure}[b]{0.5\linewidth}
    \centering
    \includegraphics[width=1\linewidth]{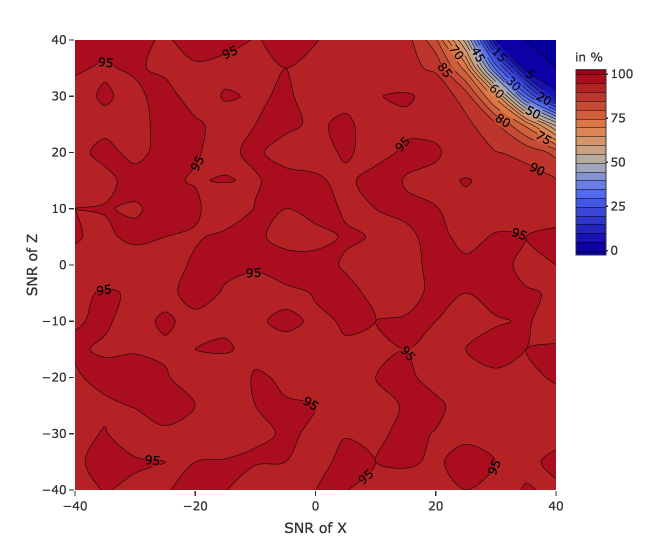} 
    \caption{\it{Rate of unidentified causality for\\ $SNR^{Y} = 40db$, in a driver topology.}} 
    \label{figb:Yback_un300D} 
    \vspace{4ex}
  \end{subfigure}
  
    \begin{subfigure}[b]{0.5\linewidth}
    \centering
    \includegraphics[width=1\linewidth]{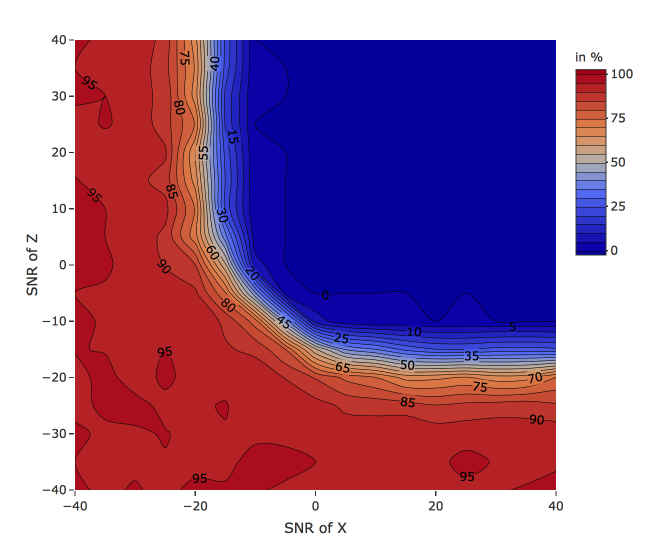} 
    \caption{\it{Rate of unidentified causality for\\ $SNR^{Y} = -40db$, in a driver topology.}} 
    \label{figc:Yfront_un300D} 
    \vspace{4ex}
  \end{subfigure}
  \begin{subfigure}[b]{0.5\linewidth}
    \centering
    \includegraphics[width=1\linewidth]{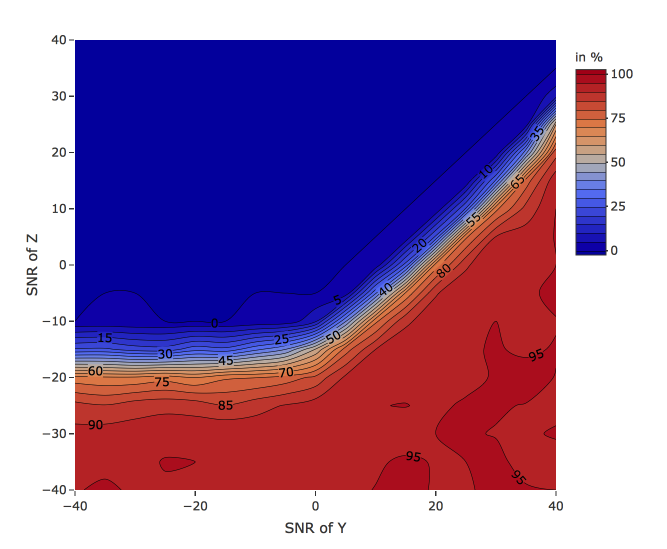} 
    \caption{\it{Rate of unidentified causality for\\ $SNR^{X} = 40db$, in a driver topology.}} 
    \label{figd:Xright_un300D} 
    \vspace{4ex}
  \end{subfigure}
  
  \caption{\it{Results for driver topology run with a sample of size of 300, encompassing the transition from from a low to high rate of unidentified causality.}}
  \label{fig:allt300unD} 
\end{figure}

\begin{figure}
  \begin{subfigure}[b]{0.5\linewidth}
    \centering
    \includegraphics[width=1\linewidth]{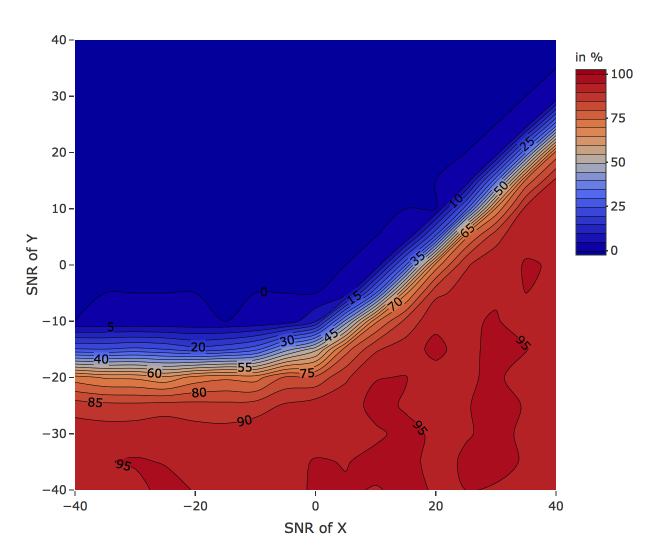} 
    \caption{\it{Rate of unidentified causality for\\ $SNR^{Z} = 40db$, in a indirect topology.}} 
    \label{figa:Ztop_un300I} 
    \vspace{4ex}
  \end{subfigure}
  \begin{subfigure}[b]{0.5\linewidth}
    \centering
    \includegraphics[width=1\linewidth]{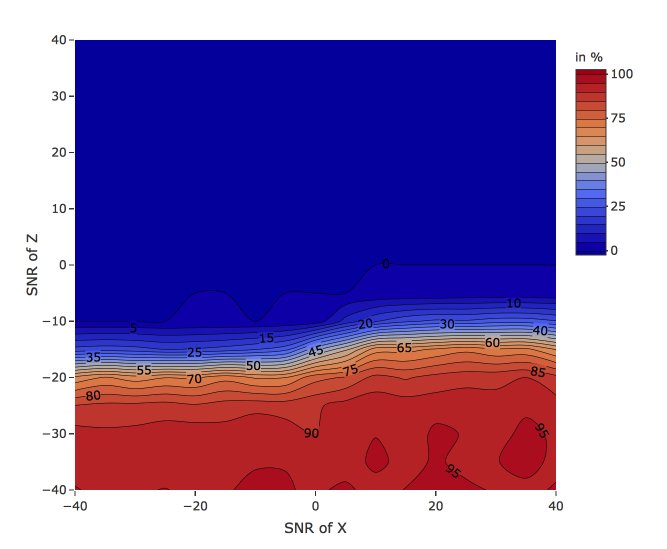} 
    \caption{\it{Rate of unidentified causality for\\ $SNR^{Y} = 40db$, in a indirect topology.}} 
    \label{figb:Yback_un300I} 
    \vspace{4ex}
  \end{subfigure}
  
    \begin{subfigure}[b]{0.5\linewidth}
    \centering
    \includegraphics[width=1\linewidth]{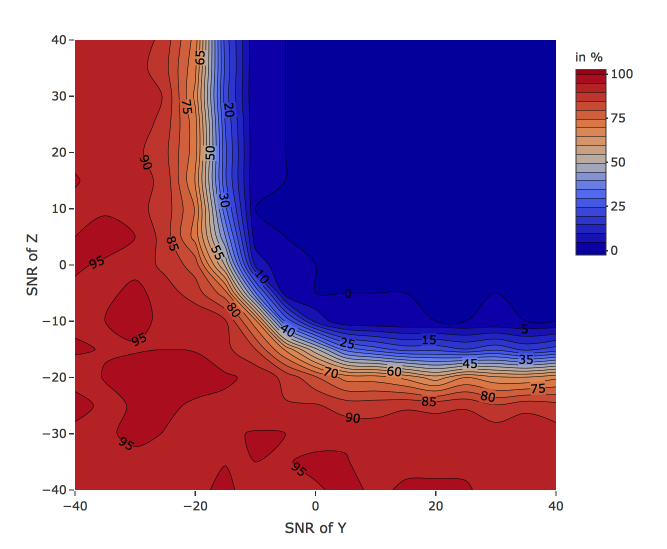} 
    \caption{\it{Rate of unidentified causality for\\ $SNR^{X} = -40db$, in a indirect topology.}} 
    \label{figc:Yfront_un300I} 
    \vspace{4ex}
  \end{subfigure}
  \begin{subfigure}[b]{0.5\linewidth}
    \centering
    \includegraphics[width=1\linewidth]{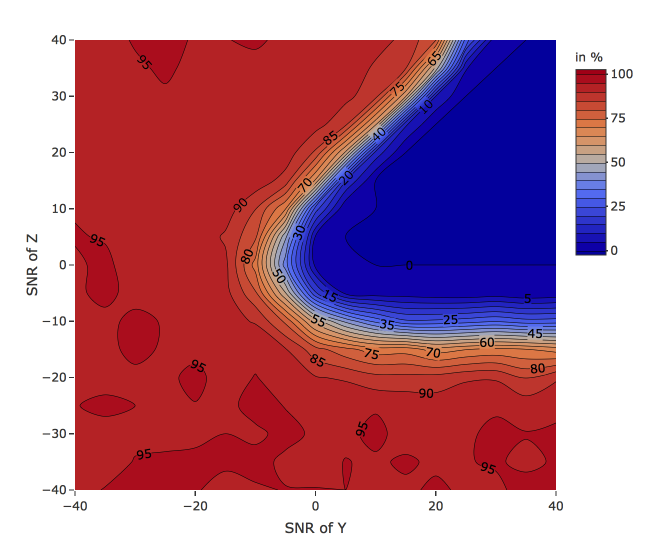} 
    \caption{\it{Rate of unidentified causality for\\ $SNR^{X} = 40db$, in a indirect topology.}} 
    \label{figd:Xright_un300I} 
    \vspace{4ex}
  \end{subfigure}
  
  \caption{\it{Results for indirect topology run with a sample of size of 300, encompassing the transition from from a low to high rate of unidentified causality.}}
  \label{fig:all300unI} 
\end{figure}

If the key link $X \rightarrow Z$ is determined to be causal this will lead to spurious causality for the indirect topology (Figure~\ref{fig:all300spI}) and if it is not determined to be causal it will lead to unidentified causality for the driver topology (Figure~\ref{fig:allt300unD}).  If the key link $Y \rightarrow Z$ is determined to be causal this will lead to spurious causality for the driver topology (Figure~\ref{fig:all300spD}) and if it is not determined to be causal it will lead to unidentified causality for the indirect topology (Figure~\ref{fig:all300unI}).

\clearpage

Figures~\ref{fig:all300spI} and \ref{fig:allt300unD} illustrate that the $X \rightarrow Z$ link is accepted as causal if both $SNR^{X}$ and $SNR^{Z}$ are sufficiently high; the threshold of acceptance depends on the value of $SNR^{Y}$ because when the quality of the $Y$ series is improved it reaches a point where it becomes as good a predictor as $X$ and hence $X\rightarrow Z$ is rejected as a casual link.
For $SNR^{Y}$ less than $0db$ if $SNR^{X}$ and $SNR^{Z}$ are greater than $0db$ then the link will always be accepted as causal, and as $SNR^{Y}$ is increased the threshold value also increases. 
Increasing $SNR^{X}$ maintains $X$'s predictive edge over $Y$ and hence the link is less likely to be rejected, but $SNR^{Z}$ must be sufficiently high
that the predictive power of $X$ can be seen (too much noise in $Z$ can mask the true signal from $X$).

Figures~\ref{fig:all300spD} and \ref{fig:all300unI} illustrate that the
$Y \rightarrow Z$ link is accepted as causal if both $SNR^{Y}$ and $SNR^{Z}$ are sufficiently high, the threshold of acceptance being dependent on $SNR^{X}$.  If $SNR^{X}$ is less then $0db$ and both $SNR^{Y}$ and $SNR^{Z}$ are greater then $-10db$ then this link is accepted. Increasing $SNR^{X}$ 
increases the threshold value.
Increasing $SNR^{Y}$ maintains $Y$'s predictive edge and hence the link is less likely to be rejected.  
As before, $SNR^{Z}$ must be sufficiently high
that the predictive power of $Y$ can be seen (too much noise in $Z$ can mask the true signal from $Y$).
Further analysis is required of the complex behaviour illustrated in Figures~\ref{fig:all300spD}(d) and \ref{fig:all300unI}(d).

Reducing the sample size to $50$ produces similar behaviour 
to that seen with intrinsic noise,
with larger spread to the transitions and a proportional change to the rates based on the significance level.

\clearpage

\subsection{Analysis}
This section has presented results for the sensitivity of Granger trivariate analysis to noise terms, investigating both intrinsic and extrinsic noise. Section~\ref{sec:intriniscNoise} and Section~\ref{sec:extrinsiNoise} demonstrate different sensitivities to the two different types of noise.

Intrinsic noise does not impact the rate of spurious causality, leading to a constant value across the phase space 
regardless of the respective $SNRs$ and regardless of the underlying topology, 
but it displays
a gradual transition from high to low unidentified causality as $SNR^Y$ and $SNR^Z$ increase, with $SNR^Y$ causing transition at a higher
value of $SNR^Z$ for a driver topology.
By contrast, extrinsic noise produces a complex non-linear transition between low and high rates for both spurious causality and unidentified causality across the phase 
space, and this transition depends not only on $SNR^Y$ and $SNR^Z$ but also on $SNR^X$.

For intrinsic noise the transition between low and high unidientified causality differs between the two underlying topologies, and for extrinsic
noise the transitions for both spurious causality and unidentified causality differ according to the key link that is being observed.
For all cases comparison between the low and high sample size versions shows that reducing sample size maintains the same behaviour while making the transition more gradual and altering the values based on the significance level. 

It should also be noted that where noise makes the rate of spurious causality high and the rate of unidentified causality low, trivariate Granger analysis is likely to return a ``complete'' topology. 
Conversely when noise makes the rate of spurious causality low and the rate of unidentified causality high, trivariate Granger analysis is likely to
return a ``null'' case (with no links) that may contradict the linkages found by bivariate Granger analysis.

\section{Summary and Conclusion} \label{sec:conclusion}

Trivariate Granger analysis seeks to improve on bivariate Granger analysis through the detection of ``spurious'' causal links that may for example be caused by driver and indirect topologies.  
Due to the extensive use of Granger causality across many fields, and the increasing interest in multivariate systems, understanding the limitations of Granger causality when applied to trivariate systems is of particular significance \citep{Eichler_2013}. 

Here we show that trivariate Granger analysis is itself capable of producing innacurate results and we analyse two causes of innacuracy:  the choice of test criteria, and the presence of noise.  Specifically, these aspects have been investigated with regard to how they affect the rates of occurrence of spurious and unidentified causality, which can both occur in either a driver or indirect trivariate topology. 

For the use of test criteria in Granger causality tests, we have expanded on work by \cite{Taylor_1989}, focusing on a trivariate topology. We have shown that the Wald and the Rao Efficient Scoring test criteria offer no statistical difference in their results, however both of these criteria offer a statistical difference to the results produced by the likelihood ratio test criterion (up to sample sizes of $75$ for spurious causality results and $25$ for unidentified causality results). These results suggest that for very low sample sizes likelihood ratio can offer a lower rate of unidentified causality, whereas Wald and Rao Efficient Scoring offer a lower rate of spurious causality. We also demonstrate that for a driver topology a far higher sample size is needed to reach a zero rate of unidentified causality, compared to an indirect topology ($300$ for the former and $175$ for the later). 

Based on work by \cite{Anderson_2018} we have investigated both intrinsic and extrinsic noise for a trivariate topology. We demonstrate that these two types of noise produce different phase spaces for both the rate of spurious and unidentified causality, showing the importance of the noise type.

In our experiments the presence of intrinsic noise did not effect the rate of spurious causality, with the entire phase space being approximately the theoretical value, i.e. the significance level of the hypothesis test. 
For the rate of unidentified causality the phase space demonstrated 
a gradual transition from high to low unidentified causality as $SNR^Y$ and $SNR^Z$ increase, with $SNR^Y$ causing transition at a higher
value of $SNR^Z$ for a driver topology ($SNR^{X}$ has no effect).

The presence of extrinsic noise in our experiments produced a complex non-linear transition between low and high rates for both spurious 
causality and unidentified causality across the phase space, and this transition depended not only on $SNR^Y$ and $SNR^Z$ but also on $SNR^X$.
We have explained how spurious causality results for the driver topology and unidentified causality results for the indirect topology both rely on the probability of the $Y \rightarrow Z$ link being detected by the trivariate Granger analysis as a causal link.  Similarly, spurious causality results for the indirect topology and unidentified causality results for the direct topology both rely on the probability of the $X \rightarrow Z$ link being detected by the trivariate Granger analysis as a causal link.  The behaviour for extrinsic noise is complex, and although we have characterised the major features of the phase space we have not yet analysed its full complexity (as displayed in Figures~\ref{fig:all300spD}(d) and \ref{fig:all300unI}(d)); this is the subject of current work, along with 
investigation of time-varying causal links as part of our broader research in the area of interaction dynamics \citep{InterDyne_1998}.

\newpage
\bibliographystyle{agsm}
\bibliography{paper_1_bib}
     
\end{document}